**PONTIFÍCIA UNIVERSIDADE CATÓLICA DE SÃO PAULO**
**PUC-SP**

Guilherme Lopes Matsushita

# Monitoramento de redes sociais – análise de ferramenta com foco no retorno estratégico empresarial

PONTIFÍCIA UNIVERSIDADE CATÓLICA DE SÃO PAULO
PUC-SP

Guilherme Lopes Matsushita

# Monitoramento de redes sociais – análise de ferramenta com foco no retorno estratégico empresarial

MESTRADO EM TECNOLOGIAS DA INTELIGÊNCIA E DESIGN DIGITAL

Dissertação apresentada à Banca Examinadora da Pontifícia Universidade Católica de São Paulo, como exigência parcial para obtenção do título de MESTRE em Tecnologias da Inteligência e Design Digital, sob a orientação do Prof. Dr. Demi Getschko.

PONTIFÍCIA UNIVERSIDADE CATÓLICA DE SÃO PAULO

PUC-SP

Guilherme Lopes Matsushita

# Monitoramento de redes sociais – análise de ferramenta com foco no retorno estratégico empresarial

MESTRADO EM TECNOLOGIAS DA INTELIGÊNCIA E DESIGN DIGITAL

Banca Examinadora

_______________________________

Orientador: Prof. Dr. Demi Getschko
Pontifícia Universidade Católica de São Paulo

_______________________________

Prof. Dr. Hermes Renato Hildebrand
Pontifícia Universidade Católica de São Paulo

_______________________________

Prof. Dr. Sergio Ricardo Master Penedo
Instituto Brasileiro de Tecnologia Avançada - IBTA

São Paulo, 06 de abril de 2016

## DEDICATÓRIA

Dedico este trabalho ao meu filho Gustavo Paci Matsushita e ao meu avô materno e padrinho, Antonio Lopes Tofetti *(in memoriam)*.

**BOLSA DE ESTUDO**

O desenvolvimento deste trabalho de pesquisa teve financiamento da CAPES.

## AGRADECIMENTOS

Agradeço à Deus pela benção e a conquista realizada de concluir um curso de mestrado, junto com todas as dificuldades encontradas e ultrapassadas.

A minha esposa, Andressa Paci Matsushita, pela dedicação, paciência, companheirismo e auxílio em todos os momentos deste período de estudos. Nos momentos fáceis e também nos difíceis, ela estava presente.

Ao meu filho Gustavo, que sempre foi e será minha inspiração e determinação para almejar sempre mais, e mostrar a importância dos estudos na vida.

Aos meus pais, Sueli e Akira, pelo sacrifício e apoio em minha caminhada acadêmica, onde os estudos sempre foram prioridade. Pela dedicação que tiveram comigo em conseguir conquistar esse título tão especial.

As minhas avós, que sempre foram exemplos de vida e espelho de sabedoria, generosidade e humildade.

Ao meu sogro Valdecir Paci e minha sogra Maria Rosa Ferreira Paci pelo apoio que me deram e pelo incentivo ao estudo.

Aos meus familiares pela cumplicidade e apoio sabendo das dificuldades de cada um, puderam ajudar mesmo a distância e também proporcionar de momentos de alegria.

Aos meus professores do TIDD, que de forma brilhante, compartilharam seus conhecimentos para enriquecer o nosso saber.

A secretária do TIDD, Edna, que sempre teve uma postura de figura orientadora aos alunos e sempre em momentos difíceis, ela se apresentava disposta a ajudar e não nos deixar desistir.

Ao meu orientador, professor Demi Getschko, que mesmo com toda turbulência do trabalho, conseguiu tempo e dedicação para a orientação que sem dúvida, foi imensurável e que, para sempre vou leva-las comigo.

Ao meu coordenador do IBTA, professor Sergio Penedo, que sempre foi figura presente a esta minha pesquisa, auxiliando e orientando nos momentos de dificuldade, e pela participação nas bancas.

Aos meus alunos, que também de forma indireta, sempre entenderam minha situação ao longo do curso, com as atividades, as provas e as orientações de TCC, e sempre me apoiaram e me deram forças para concluir este trabalho, como um exemplo para todos.

Agradeço aos colegas de estudo do TIDD, que também fizeram parte dessa história, em todos os eventos, palestras e também nas aulas, compartilhando as dificuldades, mostrando os atalhos, e sempre apoiando uns aos outros.

Agradecer a CAPES, pelo fornecimento da bolsa de estudo, que sem dúvida, foi determinante para a conclusão da pesquisa e de poder trazer este trabalho como retorno de todo o investimento.

O pessimista vê dificuldade em cada oportunidade; o otimista vê oportunidade em cada dificuldade.

**Winston Churchill**

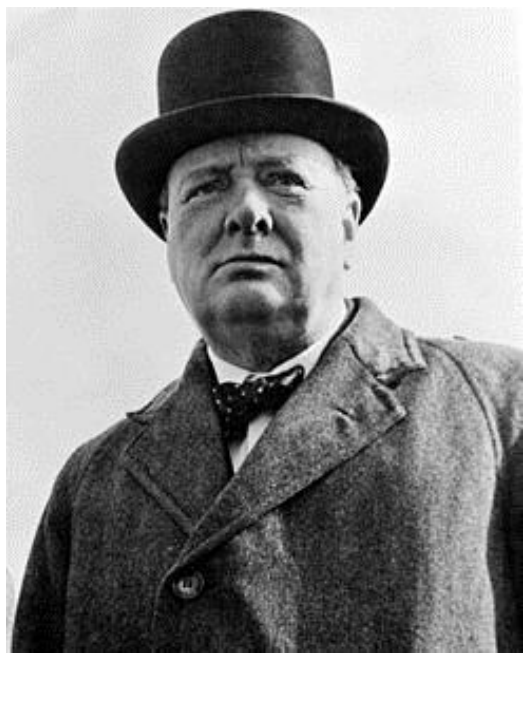

# RESUMO

Com o surgimento recente e a explosão de usuários nas redes sociais, os internautas passaram a conhecer mais sobre outras pessoas ou empresas por meio de informações obtidas de grupos ou comunidades on-line e assim conhecer novas oportunidades. Uma avalanche de informações diversas e não estruturadas é criada minuto a minuto, por bilhões de dispositivos conectados na Internet à todo o momento. Analisar, processar, compreender e avaliar esses dados, é uma tarefa desafiadora, levando em conta o volume, a velocidade e a variedade das informações obtidas nessas redes sociais. Porém, uma melhor compreensão do processo de monitoramento das redes e a utilização de ferramentas desta natureza, são essenciais para as pessoas, as empresas e para o universo acadêmico. A aplicação de sistemas de monitoramento, a gestão da informação e a possibilidade de aproximar marcas às pessoas, são alguns pontos positivos que esta pesquisa irá apresentar conhecendo melhor a problemática. Nesta pesquisa será analisada a ferramenta de monitoramento de redes sociais *SCUP*, seus detalhes, suas funcionalidades e a relação da ferramenta com a gestão empresarial. Como conclusão, são apresentadas todas as particularidades do software e as possibilidades aplicáveis no ambiente corporativo, como um auxílio às tomadas de decisão.

## ABSTRACT

With the recent emergence and the explosion of users in social networks, Internet users have come to know more about other people or companies through information obtained from online groups or communities and therefore meet new opportunities. An avalanche of diverse and unstructured information is created every minute, for billions of devices connected to the Internet all the time. Analyze, process, understand and evaluate these data, it’s a challenging task, given the volume, velocity and variety of information from these social networks. However, a better understanding of the process of monitoring networks and the use of tools of this kind are essential for people, companies and the academic world. The application of monitoring systems, information management and the possibility of bringing brands to people, are some positives that this research will provide better knowledge of the problems. In this research will be analyzed monitoring tool of social networks SCUP, its details, its features and the tool related to business management. In conclusion, we present all the peculiarities of the software and the possibilities applicable in the corporate environment, as aid to decision making.

## LISTA DE FIGURAS

## LISTA DE TABELAS

## LISTA DE SIGLAS E ABREVIATURAS

**ABCiber** Associação Brasileira de Pesquisadores em Cibercultura

**AFP** L'Agence France-Presse

**API** Application Programming Interface (em português, Interface de Aplicação de Programação)

**BBC** British Broadcasting Corporation

**CRM** Customer Relationship Management (em português, Gerenciamento de Relacionamento com o Cliente)

**DNA** Deoxyribonucleic Acid (em português, Ácido Desoxirribonucleico)

**EUA** Estados Unidos da América

**HTML** HyperText Markup Language (em português, Linguagem de Marcação de Hipertexto)

**HTTP** Hypertext Transfer Protocol (em português, protocolo de transferência de hipertexto)

**iOS** iPhone Operational System

**JSON** Javascript Object Notation

**KPI** Key Performance Indicator (em português, indicador-chave de desempenho)

**ROI** Return of Investiment (em português, Retorno do Investimento)

**RSS** Really Simple Syndication

**SAC** Serviço de Atendimento ao Consumidor

**SLA** Service Level Agreement (em português, Acordo de Nível de Serviço)

**SMS** Short Message Service (em português, Serviço de Mensagens Curtas)

**SXSW** South by Southwest

**TAM** Taxi Aéreo Marília (atual empresa TAM Linhas Aéreas)

**TIC** Tecnologia da Informação e Comunicação

| | |
|---|---|
| **TV** | Televisão |
| **URL** | Uniform Resource Locator (em português, Localizador Padrão de Recursos) |
| **WEB** | World Wide Web (ou simplesmente Web) |
| **XML** | eXtensible Markup Language |

## SUMÁRIO

# Capítulo I

## Introdução

# INTRODUÇÃO

O mundo está conectado e interligado através de redes virtuais onde a informação e o conhecimento dominam o ambiente e cada vez mais o compartilhamento fez tanto sentido quanto o que estamos vivenciando nessa nova era, a revolução tecnológica.

A informação como suporte ao processo de tomada de decisão corporativa é um assunto estudado em livros de administração, gestão empresarial entre outros, todavia, a tomada de decisão baseada somente na experiência própria do gestor se torna uma necessidade de agir com rapidez e precisão.

Não é aceitável nos dias de hoje que uma empresa que almeja a competitividade com vantagens, sem a utilização de ferramentas tecnológicas. Estes episódios abrem espaços para que os futuros gestores, com novas visões empresariais, busquem o melhoramento contínuo para suas organizações.

Descobrir a informação exata e apropriada pode ser um incômodo frente à quantidade de dados que são disponibilizados nos mais variados meios de encontrar a informação pontual e adequada podendo ser um problema frente à quantidade de sistemas computacionais que gerenciam toda essa informação para auxiliar nos processos manuais de uma empresa.

As mudanças decorrentes do desenvolvimento tecnológico nas áreas da informação e telecomunicação atingiram significativamente a sociedade. Para seguir todas essas transformações, tanto a população quanto as empresas, têm procurado por formas mais ágeis para se inserirem nesse modelo atual de mercado globalizado.

Modelo esse chamado de “era da informação”[1], a qual é preciso ter a noção da tecnologia da informação e os sistemas computadorizados como os grandes

[1] Nome dado ao período pós-industrial da sociedade moderna, mais especificamente depois da década de 80; http://mundoeducacao.bol.uol.com.br/geografia/era-informacao.htm;

antecessores e responsáveis pelo valor acrescido aos processos de tomada de decisão empresarial.

A presente pesquisa irá tratar sobre o monitoramento das redes sociais, dando ênfase à ferramenta SCUP[2], um software desenvolvido para monitorar, gerenciar e quantificar informações sobre um determinado termo pesquisado. Desta forma, podemos entender seu mecanismo e os processos para o entendimento dos resultados obtidos, discutir a forma como elas podem auxiliar a gestão corporativa atual e finalizar pontuando a influência dessas ferramentas no cotidiano empresarial.

Algumas questões foram levantadas para se buscar respostas ou até propostas de melhorias ou sugestões de alternativas ao tema proposto. Desta forma devemos destacar alguns pontos críticos a serem debatidos, são eles:

- Como funcionam as ferramentas de monitoramento das redes sociais?
- Por que monitorar as redes sociais?
- Qual a consequência do monitoramento para a gestão empresarial?

Esses são alguns pontos que o trabalho irá tentar recortar para o desdobramento da pesquisa, incluindo uma análise da ferramenta de tecnologia de monitoramento em mídias sociais para analisar informações, fazer a coleta, filtrar e interpretar dados alguma pessoa ou empresa. Esta análise inclui o estudo da solução de mercado SCUP.

Sendo assim, este trabalho visa colaborar de forma didática e acadêmica, um setor que a tecnologia permitiu alcançar através do desenvolvimento de softwares capazes de fazer todo um trabalho analítico e quantitativo de indexação de dados não estruturados e conseguir transformá-los em informações entendíveis para os negócios.

---

[2] SCUP é uma plataforma de monitoramento de mídias sociais utilizado por mais de 24 mil profissionais. Foi fundada em 2009 na cidade de São Paulo, e propaga como missão e objetivo de negócio, ajudar empresas a trabalhar de forma eficiente com as redes sociais, facilitando a rotina de quem trabalha na operação e eficiência no monitoramento, desde a coleta dos dados até a análise do resultado. (https://www.scup.com/pt/)

## 1.1 Introdução

Atualmente, a Internet surge como um ambiente democrático no qual a informação pode ser gerada, armazenada, distribuída e coletada de diversas maneiras. Novas ferramentas possibilitam que usuários criem gratuitamente seus próprios conteúdos digitais, o que contribui para o aumento da quantidade de informações disponíveis. Esse crescimento exige que torne cada vez mais necessário o uso mecanismos eficazes e eficientes para recuperação de conhecimentos na Web.

O espaço empresarial está se alterando ininterruptamente, tornando-se mais difícil e menos previsível, e cada vez mais condicionado de informações e toda a infraestrutura de tecnologia que garante a gestão de grandes volumes de dados.

## 1.2 Estado da arte

Alguns autores já trataram sobre o assunto de monitoramento das mídias sociais, e tiveram destaques em várias frentes como marketing, negócios, tecnologia entre outras. Vejamos o que já foi pesquisado e discutido sobre o tema.

FREIRE, C.P. (2014) em sua tese de doutorado, diz ter feito uso da ferramenta SCUP em campanhas políticas de 2010, como forma de monitoramento dos termos relacionados com os pré-candidatos à prefeitura de São Paulo. Em sequência, a autora também destaca que a ferramenta auxiliava em processos de classificação e customização das informações:

> A atividade consistia, basicamente, em análise e classificação de dados coletados automaticamente pelo SCUP. A técnica permitia desenvolver categorias customizadas aplicadas a conteúdos e comportamentos de usuários e, nesse sentido, a classificação proposta para usuários nunca deixou de ser algo interessante. (FREIRE, C.P, 2010, p.104)

ANTUNES, M.F. et al, em seu artigo “Monitoramento de informação em mídias sociais: o e-Monitor Dengue”, fez menção sobre o processo de monitoramento de informações na internet e nas mídias sociais, sendo assim, a relação com esta pesquisa fica mais estreita no sentido do monitoramento de palavras-chave:

> Este programa percorre a Internet recolhendo informações, que são indexadas em uma base de dados e consultadas posteriormente por mecanismos de busca. Assim, o programa-robô é o responsável por monitorar as mudanças nas fontes cadastradas no sistema de monitoramento, emitindo um alerta sempre que um novo conteúdo for publicado. Desta forma, o sistema de monitoramento é capaz de monitorar e capturar informações, por dados pré-definidos (palavras-chave, fontes e links). (ANTUNES, M.F. et al, p.9-18, jan./abr., 2014)

A análise das redes sociais pode contribuir os processos de tomada de decisão estratégica corporativa, além de ajudar a compreensão da estrutura informacional da organização. Esses pontos foram estudados por ALCARÁ, A.R. et. al, em seu artigo chamado “As redes sociais como instrumento estratégico para a inteligência competitiva”. A autora destaca para a identificação dos fluxos de compartilhamento da informação.

> A análise de redes sociais, além de contribuir para a compreensão do organograma institucional e estrutura de recursos humanos, é importante para o entendimento da estrutura informacional da organização, visto que permite a identificação dos fluxos de compartilhamento da informação e do conhecimento. (ALCARÁ, A.R. et al, TransInformação, Campinas, 18(2):143-153, maio/ago., 2006)

Outros trabalhos relacionados ao tema também fazem parte desta pesquisa sobre as redes sociais e a coleta dos dados dos usuários através de aplicações específicas. SOUZA, F. B. e colaboradores citando NAZIR, em seu artigo dizem:

> Aplicações podem ser utilizadas para o estudo de interações entre os usuários que utilizam as aplicações e também podem ser úteis para coletar outras informações dos usuários, tais como lista de amigos e atividades executadas durante uma sessão. Alguns trabalhos fizeram uso dessa estratégia para estudar os usuários de redes sociais online. Nazir e colaboradores analisaram características de aplicações no *Facebook*, desenvolvendo e lançando suas próprias aplicações. Em particular, eles estudaram a formação de comunidades online a partir

> de grafos de interação entre os usuários de suas aplicações. Mais recentemente, os mesmos autores estudaram várias características relacionadas ao desempenho de suas aplicações no *Facebook*. (SOUZA, F.B., et al, 2010)

## 1.3 Justificativa

A forma como os negócios são gerenciados, e são adaptados com novas ferramentas tecnológicas, sempre foi uma questão de mudança de paradigma para muitas organizações. Empresas do mundo todo se encontram num patamar tecnológico onde, quem detém estratégias de mercado bem definidas utilizando dessas ferramentas e recursos, consegue alcançar num período menor os objetivos propostos pela organização.

Isso ocorre, pois consegue de forma estruturada, melhor se organizarem, conseguindo resultados claros e diretos, otimizando e elevando as tomadas de decisão para a alta administração, baseando-se em informações reais, objetivas e verdadeiras.

A tecnologia das redes mundiais de computadores, por sua vez, tem papel fundamental para as empresas, com tal importância que não é possível utilizá-la somente, e sim, compreendê-la, e se sujeitar às suas mudanças, que a todo o momento surgem em algum lugar do mundo.

As redes sociais, as mídias da internet de forma geral, como vídeos, blogs, fotos, sites e fóruns, podem ser os novos canais da expressão pública em massa, que a sociedade moderna e digital poderá fazer proveito para um benefício comum.

Para melhor aproveitar toda a inteligência coletiva disponível na Web, empresas utilizam algumas “ferramentas analíticas” [3] de interpretação de dados estruturados, na forma de leitura dinâmica das informações geradas por toda a internet. Essas ferramentas atualmente servem para observar as preferências, o comportamento, bem como expressão da opinião do consumidor na Web.

---

[3] Ferramentas analíticas – *Google Analytics; Facebook Insights; Twitter Analytics; Social Crawlytics* entre outras.

O excesso de informação gerada e compartilhada nas redes sociais a respeito das empresas e sobre seus produtos e serviços, constitui em um problema de gestão estratégica empresarial. Qual o impacto das informações anunciadas nas mídias sociais, para que o gestor possa tomar suas decisões precisas e rápidas, tendo o conhecimento prévio do que está sendo discutido e comentado na rede virtual, no momento da sua decisão? Essa é uma questão levantada diante do cenário atual na internet.

Os paradigmas da governança corporativa nos padrões atuais podem ser remodelados para que tenhamos uma performance[4] do uso de ferramentas desenvolvidas com mais recursos para a integração dos dados das redes sociais, em seus sistemas de gestão atuais, para nos preparar para o futuro, diante de milhares de informações oriundas de várias formas e lugares diferentes, ao ponto de saber informar, de modo inteligente e estratégico, quais informações são importantes e relevantes sobre a empresa, para a tomada de decisão na gestão corporativa.

Saber como as informações disponibilizadas afetam a gestão, ao ponto de corrigir um problema desconhecido, ou alterar um processo desorganizado, um funcionário trabalhando de forma a prejudicar a empresa, é ponto chave que o gestor deve considerar na tomada de decisão, sendo as mídias sociais, uma fonte de dados em tempo real e gratuita.

Consumidores, clientes, fornecedores ou até mesmos os próprios colaboradores podem contribuir, para uma melhor gestão empresarial ou colaborar construtivamente, nas mídias sociais, com informações desconhecidas ao gestor, mas que podem fazer grande diferença aos que utilizam os serviços ou compram os produtos.

---

[4] Neste caso, performance tem sentido de melhoria, relacionado à processos.

Descobrir a melhor estratégia de lidar com essa "tonelada de informações", chamada de "*big data*"[5] é uma das maneiras de como as empresas irão gerenciar as informações internas com a colaboração externa dos usuários.

Como a velocidade, a variedade e o volume dos dados das mídias sociais digitais podem influenciar a maneira de gestores tomarem decisões estratégicas ou as empresas alterarem o seu modo de gerenciar os processos internos, será a motivação intrínseca dessa pesquisa.

## 1.4 Objetivos

### 1.4.1 Objetivo geral

Analisar o processo de monitoramento das redes sociais de acordo com a ferramenta SCUP, suas funcionalidades e benefícios voltados à gestão estratégica empresarial.

### 1.4.2 Objetivos específicos

- Definir o conceito de redes sociais digitais;
- Analisar todo o processo de varredura das redes sociais (de forma genérica) para explicitar todas as etapas internas da ferramenta;
- Entender o padrão das métricas utilizadas para mensurar as informações na ferramenta de monitoramento;
- Entender o processo de análise do conteúdo obtido através da busca e monitoramento de palavras-chave inseridas na ferramenta.

---

[5] Big Data é o termo que descreve o imenso volume de dados – estruturados e não estruturados – que impactam os negócios no dia a dia. Mas o importante não é a quantidade de dados. E sim o que as empresas fazem com os dados que realmente importam. Big Data pode ser analisado para a obtenção de insights que levam a melhores decisões e direções estratégicas de negócio. (http://www.sas.com/pt_br/insights/big-data/what-is-big-data.html)

## 1.5 Hipóteses

Com a disponibilidade do uso de ferramentas de monitoramento de redes sociais na Internet, é necessário avaliar o retorno da implementação dessa solução no ambiente empresarial e garantir um ganho de performance dos resultados da gestão estratégica.

Como hipótese desta pesquisa, teremos um diagnóstico sobre as mudanças ocorridas no processo de gestão empresarial, após a utilização da ferramenta de monitoramento das redes sociais, como suporte à gestão de relacionamento com o cliente, ouvidoria da empresa, e canal de marketing e relacionamento direto.

## 1.6 Delimitação da pesquisa

Essa pesquisa, de acordo com a área estudada, irá tratar de assuntos pertinentes aos conceitos macros sobre mídias sociais, plataformas colaborativas na Internet e os objetos de estudo de caso apontados no trabalho.

Também faz parte do campo de atuação, a leitura de autores da área das mídias digitais como referências bibliográficas, artigos e trabalhos acadêmicos sobre a utilização de ferramentas de monitoramento das redes sociais no âmbito corporativo.

Deste modo, esta pesquisa não irá adentrar no campo comportamental, psicológico ou afetivo do gestor ou de outros envolvidos no processo de tomada de decisão. Assuntos desta natureza não serão especificados ou detalhados neste trabalho, pois não farão parte da temática central da pesquisa.

## 1.7 Fundamentação teórica

As mídias sociais são os elementos principais desta pesquisa, e fundamentais para a compreensão sobre o quão integrado elas estão em nosso meio e como essa relação afeta a nossa sociedade.

Para dar início ao entendimento sobre as mídias sociais e todo o conjunto relacionado neste trabalho, devemos entender a visão sobre as redes interativas citadas por Pierre Levy (2013):

> As redes interativas de computadores estão crescendo exponencialmente, criando novas formas e canais de comunicação, moldando a vida e, ao mesmo tempo, sendo moldada por elas. As mudanças sociais são tão drásticas quanto os processos de transformação tecnológica e econômica. (LÉVY, P. 2013)

Pierre Levy percebe a importância em relação às mudanças que as redes de computadores causam em nosso ambiente.

Quando analisamos redes sociais, temos a versão de Raquel Recuero (2010) que diz:

> Uma rede social é definida como um conjunto de dois elementos: atores (pessoas, instituições ou grupos; os nós da rede) e suas conexões (interações ou laços sociais). Uma rede, assim, é uma metáfora para observar os padrões de conexão de um grupo social, a partir das conexões estabelecidas entre diversos atores. (RECUERO, R. 2010)

Como complemento desta citação, temos a versão de Laura Garton[6] (2006), onde sintetiza o conceito de rede social da seguinte forma: “Quando uma rede de computadores se conecta com pessoas ou organizações, temos uma rede social”.

Esses conceitos sobre redes interativas e também sociais estão alinhados pelos seus elementos, que juntos formam uma nova dimensão.

---

[6] http://onlinelibrary.wiley.com/doi/10.1111/j.1083-6101.1997.tb00062.x/full

Desta forma, as comunidades virtuais chegaram com um propósito social do ciberespaço, e sendo assim parte integrante para o futuro da sociedade digital.

Para somar a estes conceitos, novamente Pierre Levy (2010) aponta sua descrição sobre a evolução das redes sociais e da sua importância para o futuro, que diz o seguinte:

> As comunidades virtuais começaram a se desenvolver há mais de vinte anos antes da aparição da Web. Hoje, elas constituem o fundamento social do ciberespaço e uma das chaves para a futura ciberdemocracia. (LÉVY, P., 2010)

Assim, a 'Rede' deixa de ser um local apenas, e passa a ter um sentido mais amplo da palavra. A internet está se transformando em um ambiente cultural que estimula a inteligência coletiva e novas formas de construir o conhecimento. Esse raciocínio é mostrado na citação abaixo de SPADARO (2013):

> Assim, a *Rede* é um local: é um ambiente comunicativo, formativo e informativo, não um "meio" "a ser usado" como um martelo ou uma antena. A internet não é um simples "instrumento" de comunicação que se pode usar ou não, mas um ambiente "cultural" que determina um estilo de pensamento a cria novos territórios e novas formas de educação, contribuindo para definir também o modo novo de estimular as inteligências e de construir o conhecimento e as relações. (SPADARO, A., 2013)

As interações e relações dos elementos das redes sociais estão evidenciadas como principal estrutura dessa dimensão digital e eletrônica que vivemos atualmente.

As redes sociais, hoje na visão de ferramentas digitais, estão cada vez mais sendo enraizada em nossas relações pessoais, de acordo com nossas preferências, grupos étnicos, políticos ou religiosos.

Os sites de relacionamentos pessoais conheceram uma nova dimensão, difundida no mundo inteiro a partir de 2004, quando um aluno de Harvard colocou na internet alguns perfis dos inscritos na universidade com o nome de *Facebook*[7]. Teve seus ideais em torno da interação de jovens estudantes, onde viram uma explosão

---

[7] *Facebook* – https://www.*Facebook*.com/

de usuários cadastrados em pouco tempo devido à sua popularidade nas redes acadêmicas.

Depois da grande divulgação pelos estudantes, a plataforma foi aberta ao público em geral a partir de 18 anos completos. Essa afirmação pode ser alinhada também com as palavras de SPARADO (2013):

> O fenômeno está incluído em um movimento mais amplo, das chamadas redes sociais, isto é, oportunidades de associação social que se expandem graças à internet. O *Facebook* se insere exatamente nessa evolução da Rede, permitindo a agregação de pessoas ligadas real ou potencialmente por algo específico (amizade, interesses...), de forma a poderem também escolher que aceitar dentro do próprio grupo de "amigos" com os quais ficar ligados. (SPADARO, A., 2013)

Outro fenômeno que aconteceu nesse recentemente, mais precisamente em 2006, foi a criação da plataforma de *microblog* chamado *Twitter*[8]. Essa ferramenta mudou a maneira das pessoas se expressarem e compartilhar informações em um novo formato, utilizando somente 140 caracteres, com aqueles que os seguem. Spadaro detalha mais sobre o *Twitter*, da seguinte forma:

> O que é o *Twitter*? Em inglês essa palavra significa "papilo", mas também pode ser um verbo, "gorjear", "chilrear". O símbolo do *Twitter* é, de fato, um passarinho que gorjeia. Trata-se de uma forma de socialização lançada na Rede em março de 2006 pela *Obvious Corporation*, da cidade americana de São Francisco. Ela permite que uma pessoa envie de um computador ou *smartphone* uma mensagem chamada *tweet*, com até 140 caracteres, que chega imediatamente àqueles que escolheram ficar em contato, seus seguidores. (SPADARO, A., 2013)

Como o sentido desta pesquisa é sobre a colaboração das redes sociais, em relação às empresas, o *Twitter* pode ser considerada uma plataforma ágil, pois assume várias frentes perante seus usuários. Continua Spadaro sobre os detalhes do *Twitter*:

---

[8] *Twitter* - https://*twitter*.com/

> Como se pode perceber, o *Twitter* é uma realidade muito flexível porque pode assumir vários significados: desde mensagem instantânea (MI), quanto de um SMS, até um verdadeiro instrumento de rede social como forma peculiar de *blog* coletivo, que permite criar, trocar e integrar ideias, notícias e conceitos; em resumo, um verdadeiro e próprio laboratório de microcomunicação em ebulição. (SPADARO, A., 2013)

Desta forma, com essas definições bem claras, é analisado o quanto as redes sociais estão integradas ao cotidiano do ser humano e sua sociedade, e as consequências geradas por essas transformações.

Pierre Lévy mostra que as integrações das redes geram um histórico da vida digital.

> “Já assistimos hoje à integração de várias redes sociais onde informações de sistemas como *Facebook*, *Linkedin*, *Twitter*, *Flicker*, *YouTube*, só para citar os mais utilizados, estão integradas a outros e, entre eles, gerando um verdadeiro “*life-log”,* ou seja, informações pessoais e comunitárias interligadas no espaço virtual global”.

As redes sociais podem exercer uma enorme força transformadora social e cultural do ciberespaço, com as comunidades virtuais e a coletividade social, pelas suas trocas de ideias e informações, concebendo um novo conceito entre os elementos, chamado de ciberdemocracia.

Lévy continua seu raciocínio citando uma analogia com os grupos sociais do mundo real. Segue os detalhes:

> “Não somente as tribos virtuais exerceram seu nomadismo nas planícies e florestas sem limites do ciberespaço, mas podemos facilmente passar de uma tribo a outra, fecundando a inteligência coletiva das comunidades virtuais pela troca e pelo entrelaçamento de seus membros e de suas ideias. É nesse novo cenário que devemos pensar e fazer surgir a ciberdemocracia”.

Partindo para a linha de pesquisa deste trabalho, relacionando as mídias sociais com a gestão estratégica empresarial, a Pesquisa sobre o uso das tecnologias da informação e comunicação no Brasil: TIC domicílios e empresas 2013, mostra a tendência das empresas estarem inseridas nas redes e utilizá-las para estratégias de marketing e vendas, já que parte das empresas conta com pessoas treinadas para estas funções.

Vamos aos fatos:

> "Em relação à presença nas redes sociais, a pesquisa mostra que 39% das empresas brasileiras que possuem acesso à Internet participam em alguma rede social, sendo que o setor de informação e comunicação possui uma adesão de 63%, enquanto o setor de alojamento e alimentação tem 51% das empresas presentes nas redes sociais. Entre as empresas que possuem perfis nas redes sociais, 66% mantêm uma área própria ou uma pessoa responsável pelo monitoramento da empresa na rede. Constatou-se também que 60% das empresas brasileiras que estão presentes nas redes sociais utilizam essas ferramentas para lançar novos produtos ou serviços, 54% para fazer promoções, e 37% para vender produtos e serviços".

Outro ponto conceitual estudado nesta pesquisa é sobre o monitoramento dos dados, portanto cabem algumas definições sobre os tópicos citados. Um deles é sobre a mineração dos dados nas redes sociais. O que Freire (2015) diz, citando KAUSHIK (2010) e STERNE (2010), é que o avanço de técnicas de análise de dados, teve como consequência uma situação chamada de "paradoxo dos dados" no sentido de que temos muitos dados e também em formas de captura, processamento e armazenamento, porém não somos capazes suficientes em gerar *insights* dessas informações.

Segundo LAINE; FRUHWIRTH (2010), algumas empresas do ramo de tecnologia em redes sociais, foram pioneiras em desenvolver metodologias de monitoramento de conteúdo dos usuários, incorporando interesses e afinidades dos perfis, a fim de comercializar essas informações para anunciantes específicos.

Desta forma se dá por encerrado os recortes bibliográficos para dar sequência ao texto, apontando em detalhes as etapas do processo de monitoramento atual das mídias sociais, com ênfase na ferramenta SCUP.

## 1.8 Estrutura do trabalho

A seguinte pesquisa está estruturada em capítulos alinhados ao tema principal e com subdivisões que dão suporte à linha de pesquisa principal.

Deste modo, seguem em detalhes os capítulos e uma breve descrição de suas abordagens:

- Capítulo 1: Introdução e estado da arte; trata da explanação inicial do tema proposto e da apresentação dos autores de referência como base bibliográfica para o decorrer do trabalho.

- Capítulo 2: Porque monitorar as redes sociais; aborda o panorama das mídias sociais, destacando o *Facebook* e o *Twitter*, como as principais redes no país, e também salientar pontos sobre o que chamamos atualmente de *"big data"*.

- Capítulo 3: Entendendo o monitoramento das mídias sociais; uma breve introdução sobre o monitoramento, seguido do processo e as métricas utilizadas para as informações obtidas e fechando o capítulo com a análise dos dados capturados nas redes sociais.

- Capítulo 4: Estudo de caso: a ferramenta SCUP; trata da história da empresa e da sua evolução, depois são apresentados todos os detalhes da ferramenta sobre o seu funcionamento e a interação com o usuário, e por fim dois casos de clientes do SCUP que utilizaram.

- Capítulo 5: Conclusões; Por fim são apresentadas as considerações finais da pesquisa e seu delineamento para as pesquisas futuras. Serão apresentadas como temáticas, o conceito de "*Big Social Data*", como projetos a serem desenvolvidos.

Desta maneira, a estrutura do trabalho mostra uma ligação entre os tópicos propostos e também uma conexão com o título apresentado, alinhando os raciocínios dos autores utilizados nas referências e o cruzamento das informações obtidas através de artigos acadêmicos e *papers* na internet.

Sendo assim, o trabalho está por completo e demarcado suas áreas de pesquisa e aprofundamento, dando sequência à metodologia.

## 1.9 Metodologia

O método utilizado está baseado na técnica de pesquisa e análise de conteúdo. Busca oferecer uma abordagem qualitativa das informações obtidas. Na pesquisa qualitativa o interesse está no processo de monitoramento das redes sociais e também na análise dos dados e suas classificações.

Portanto, é uma pesquisa do tipo exploratória, que não tem a ambição de chegar a resultados conclusivos e passíveis de generalização, mas sim colaborar para a formação de novos conhecimentos ao redor do assunto proposto, que segundo Freitas e Castro (2004, p.3): “O resultado que se busca é a compreensão e descrição do processo, não respostas tidas como verdades, levando-se em conta que as representações sociais estão em constante reformulação”. Assim, o fato analisado é temporário e aberto, devendo ser questionado e mais aprofundado.

# Capítulo II
# Porque monitorar as redes sociais

# 2 PORQUE MONITORAR AS REDES SOCIAIS

As redes sociais digitais estão repletas de fontes de informações e dados preciosos que podem ser analisados e utilizados para fins de estratégia competitiva no mercado empresarial. Este capítulo visa elaborar uma visão geral sobre as mídias sociais no cenário brasileiro e também um estudo sobre as questões acerca do grande volume de dados na Internet.

## 2.1 O universo digital das mídias sociais

A velocidade com que as redes sociais na Internet conquistaram adeptos é, no mínimo, assustadora. Elas são *sites*[9] construídas para interação entre pessoas, combinando textos, imagens, sons e vídeos.

A tabela 1 mostra uma lista das principais redes sociais na atualidade e seus propósitos.

**Tabela 1 – Principais redes sociais na internet**

| Nome | Propósito | Endereço Web |
|---|---|---|
| ***Facebook*** | Relacionamento e amizade | https://www.*Facebook*.com/ |
| ***Twitter*** | Relacionamento e amizade | https://*Twitter*.com/ |
| **Linkedin** | Contatos profissionais | https://br.linkedin.com/hp/ |
| **Instagram** | Compartilhamento de foto e vídeo | https://www.instagram.com/ |
| **Pinterest** | Compartilhamento de foto | https://br.pinterest.com/ |
| ***YouTube*** | Compartilhamento de vídeo | https://www.*YouTube*.com/ |
| **Google+** | Relacionamento e identidade | https://plus.google.com/ |
| **Tumblr** | Compartilhamento multimídia | https://www.tumblr.com/ |
| **Alvanista** | Relacionamento e jogos | http://alvanista.com/ |
| **Last.fm** | Relacionamento e música | http://www.last.fm/pt/ |
| **Qzone** | Relacionamento e amizade | http://qzone.qq.com/ |

(Fonte: elaborada pelo autor)

[9] Site (em português, sítio) é um conjunto de páginas web, desenvolvidas em linguagem estruturada, acessadas através de um protocolo por um navegador web.

No Brasil, mais de 80% dos internautas participam de alguma dessas mídias, segundo pesquisa[10] realizada pela empresa comScore[11] e divulgada em setembro de 2011: Orkut, *Twitter*, *MySpace*, *YouTube*, *Facebook*, *Flickr*, LinkedIn e tantos outros são usados para a troca de mensagens, compartilhamento de interesses, atualização de perfis. Informação do mundo para o mundo. Bem diferente de alguns anos atrás, quando a comunicação de massa era de um para todos: da TV para o telespectador, do jornal para o leitor, do rádio para o ouvinte.

O universo das redes sociais está em constante expansão e o mundo dos negócios responde através do ritmo acelerado de adoção de estratégias para gerar conteúdo online, melhorar o atendimento ao cliente e aumentar sua notoriedade.

> Na era das mídias sociais, branding é o diálogo que você tem com seus atuais e potenciais clientes. Quanto maior o diálogo, mais forte a marca; quanto mais fraco o diálogo, mais fraca a marca. Graças à internet, esse diálogo pode acontecer 24 horas por dia, 365 dias por ano. Isso inclui tanto a conversação que você mantém com seus clientes quanto às trocas que eles mantêm entre si - tudo representa o fortalecimento da marca. (WEBER, 2007, p.19)

Atualmente os números representativos sobre a internet e as mídias sociais no Brasil, possuem números expressivos diante dos números globais, como mostrado na figura 1. Outra informação que merece destaque são as redes sociais mais utilizadas e acessadas pelos usuários brasileiros, de acordo com a pesquisa realizada em 2015; conforme a figura 2.

---

[10] http://www.comscore.com/por/Imprensa-e-eventos/Apresentacoes-e-documentos/2011/The-Rise-of-Social-Networking-in-Latin-America

[11] A comScore é uma companhia líder em tecnologia de internet que mede o que as pessoas fazem enquanto elas navegam pelo universo digital – e transforma essa informação e insights e ações para que nossos clientes maximizem o valor de seu investimento digital.

**Figura 1 – Principais indicadores digitais do Brasil**

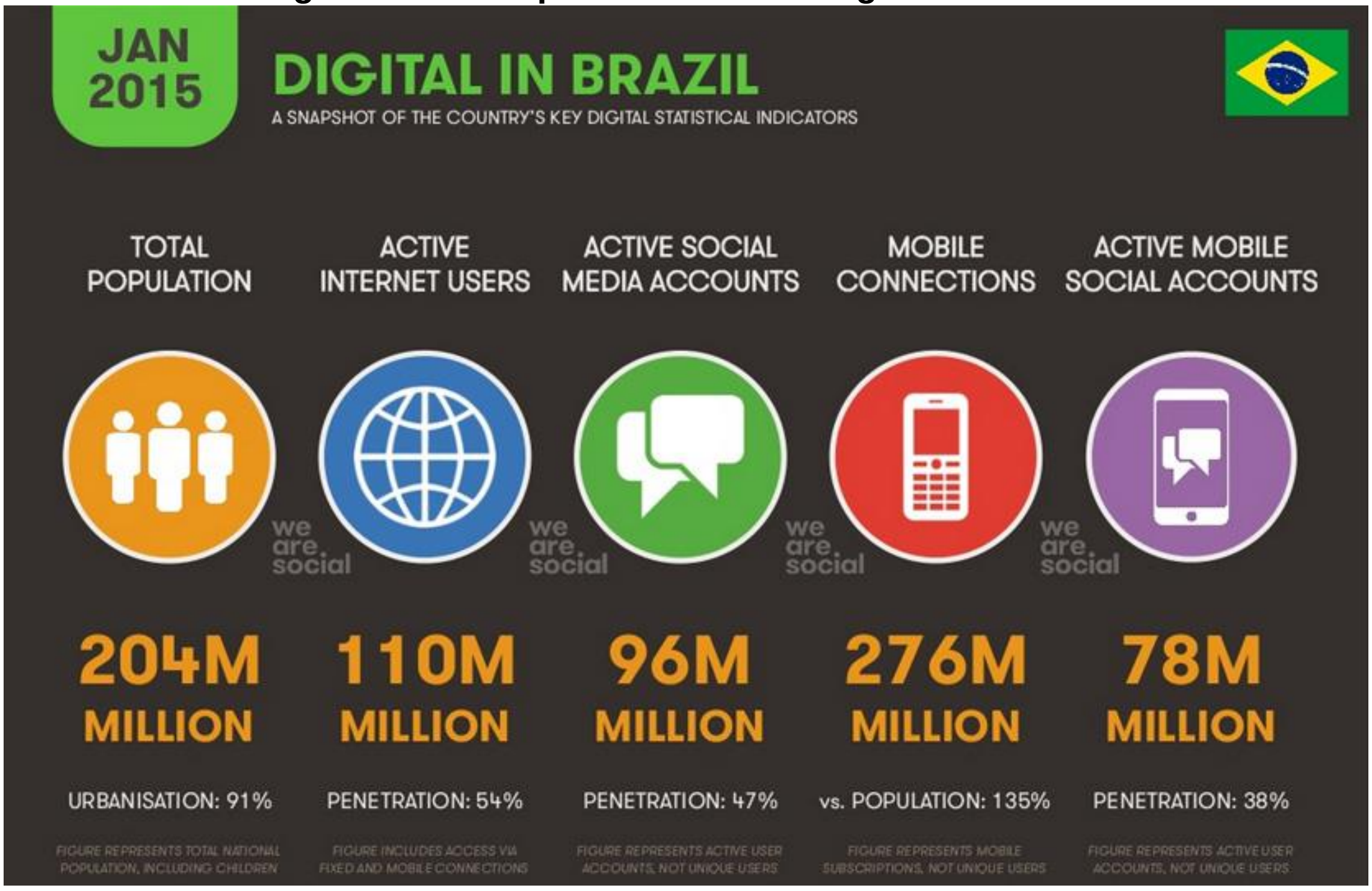

(Fonte: http://blog.pmweb.com.br/a-internet-no-brasil-em-2015/)

**Figura 2 – As plataformas sociais ativas mais utilizadas no Brasil**

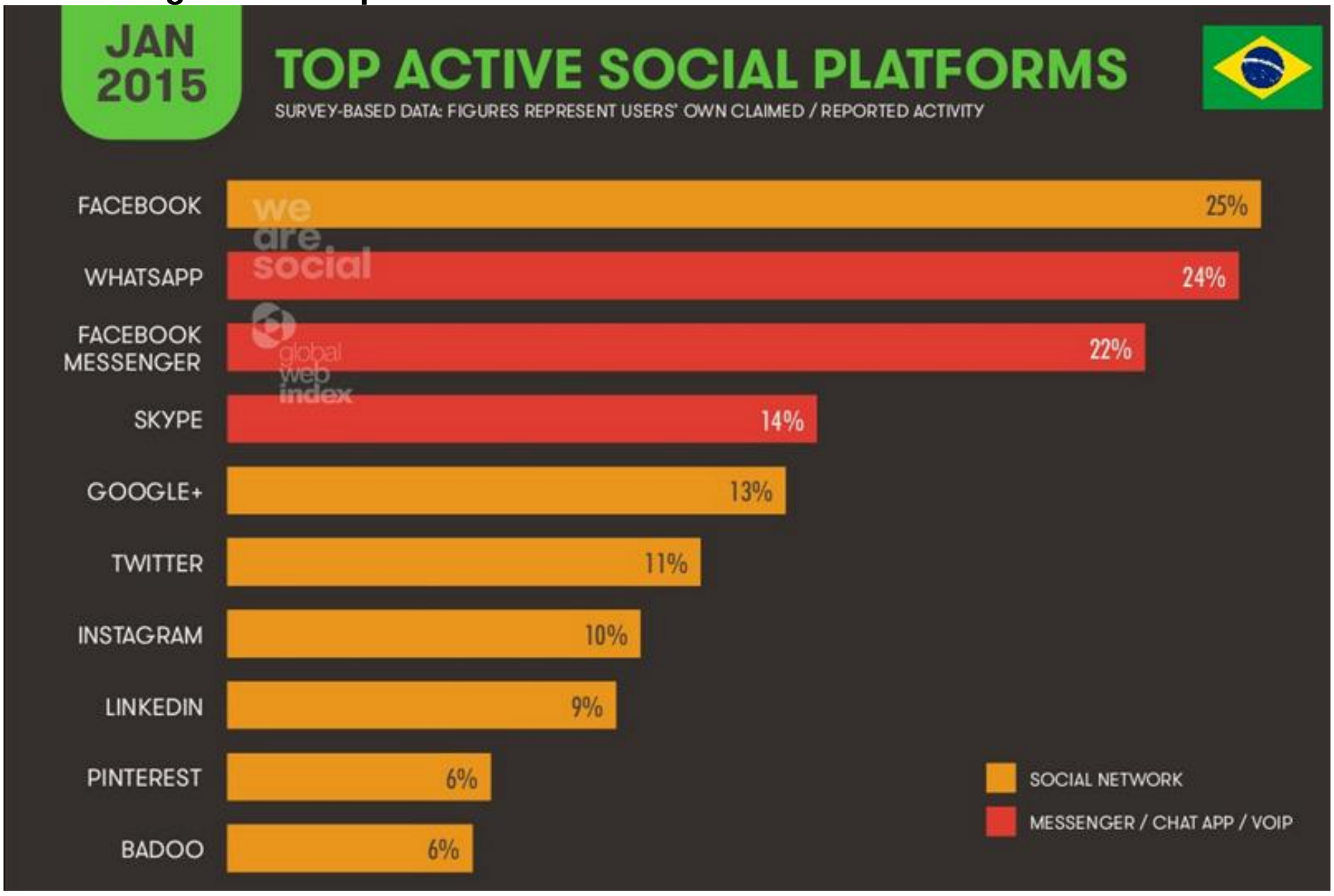

(Fonte: http://blog.pmweb.com.br/a-internet-no-brasil-em-2015/)

O universo das mídias sociais está em pleno crescimento, onde a cada momento nasce uma nova rede especializada para um perfil de usuário ou grupo social. Neste sentido, vale ressaltar duas das principais redes sociais na internet e seus impressionantes números.

## 2.2 O *Facebook* e o *Twitter*

As duas redes sociais mais populares dos últimos anos são os objetos de estudo dos tópicos a seguir.

### 2.2.1 *Facebook*

O *Facebook* é uma rede social digital que admite amigos conversarem e compartilharem mensagens, fotos, vídeos e conteúdos diversos. Foi fundado em 2004 na universidade americana de Harvard, por três amigos, sendo o mais conhecido Mark Zuckerberg e a coparticipação do brasileiro Eduardo Saverin, com o intuito de socializar o campus da universidade através de uma rede de amizade, informando quem era solteiro ou comprometido, aumentando a popularidade do indivíduo dentro do campus.

A figura 3 mostra o visual da primeira página na Web da rede social, bem simples e de *layout* rústico. Sem muitas opções, ela mostrava o perfil do usuário e suas principais informações e seu foco era mesmo no universo acadêmico.

**Figura 3 – Página inicial do *Facebook* (primeira versão)**

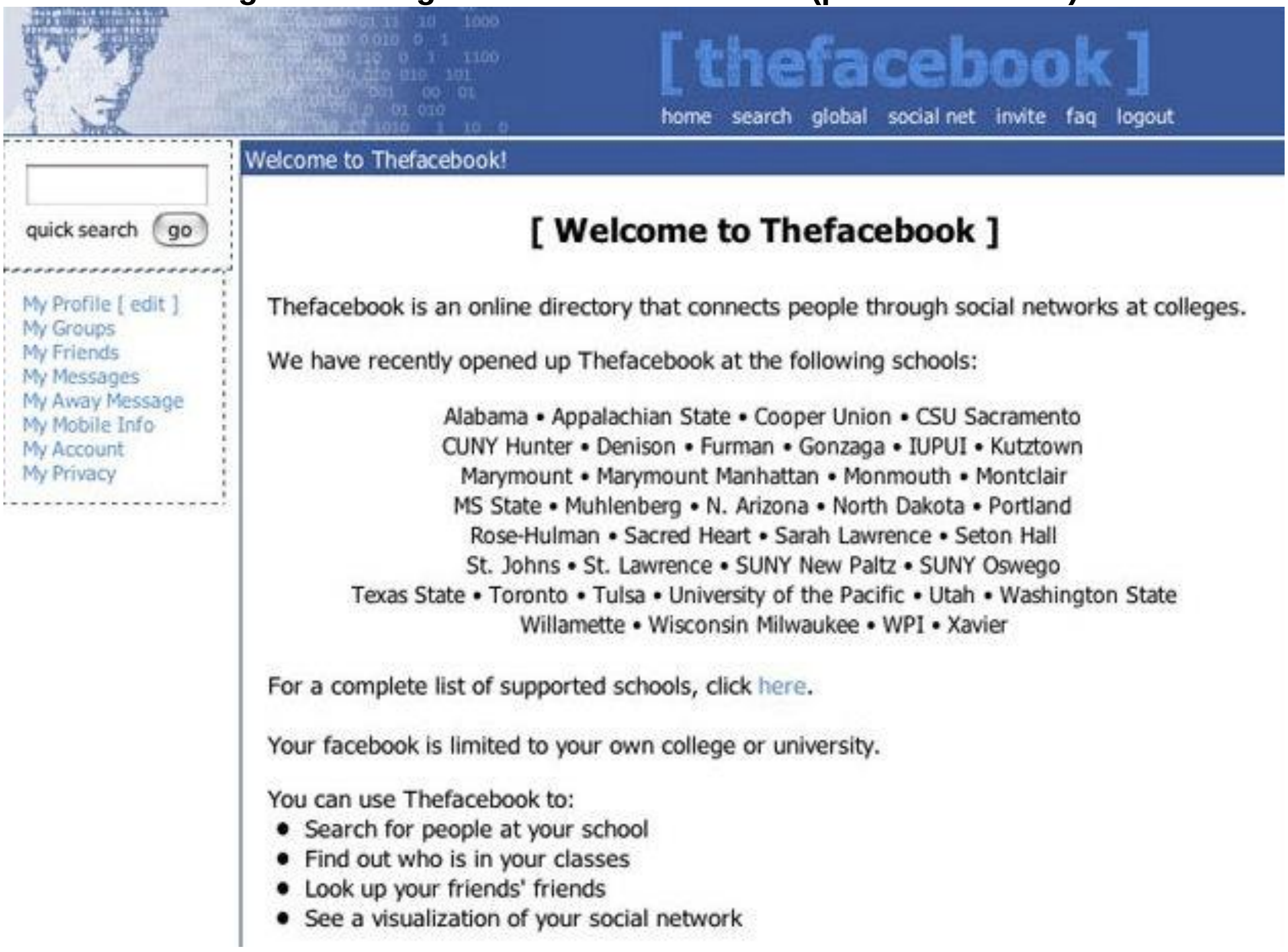

(Fonte: http://files.itproportal.com/wp-content/uploads/photos/the-*Facebook*-1_original.jpg)

Com todo o sucesso na universidade de Harvard, a página se espalhou para outras faculdades dos Estados Unidos e seu nome mudou de "*The Facebook*" para somente "*Facebook*", onde alcançou o extraordinário número de 5,5 milhões de membros no ano de 2005.

Com o intuito de angariar mais usuários, o *Facebook* começou a cadastrar pessoas com mais de 13 anos, mesmo não tendo vínculo com uma instituição de ensino. Sendo assim, no ano de 2006 já contabilizava cerca de 12 milhões de usuários, o que fez com que o *layout* ficasse parecido com o atual. Foi adotada a barra azul no topo das páginas com o fundo branco e também foi criado um "*feed"* (um tipo de lista de atualizações dos usuários e das páginas do *Facebook*) pessoal na página de cada usuário, apresentando tudo o que era publicado. Essas alterações estéticas fizeram com que a rede conquistasse mais usuários.

Foi em meados de 2007 que os aplicativos começaram a ser integrado ao *Facebook*, o que permitiu outro tipo de interação com os amigos virtuais. A partir desta implementação que surgiram fenômenos como "*Farmville*", um dos jogos de maior sucesso, e "*Mafia Wars*", que permitia compartilhar as conquistas ou solicitar ajuda aos amigos da rede social, caracterizando um tipo de colaboração social.

No ano de 2008, o *Facebook* passou o concorrente *MySpace* e se tornou a maior rede social do mundo em números de usuários. Em outubro do mesmo ano, já era contabilizado 100 milhões de pessoas conectadas. A figura 4 mostra a página inicial da rede neste ano. A página era mais moderna e com vários recursos disponíveis.

**Figura 4 – Página inicial do *Facebook* em 2008**

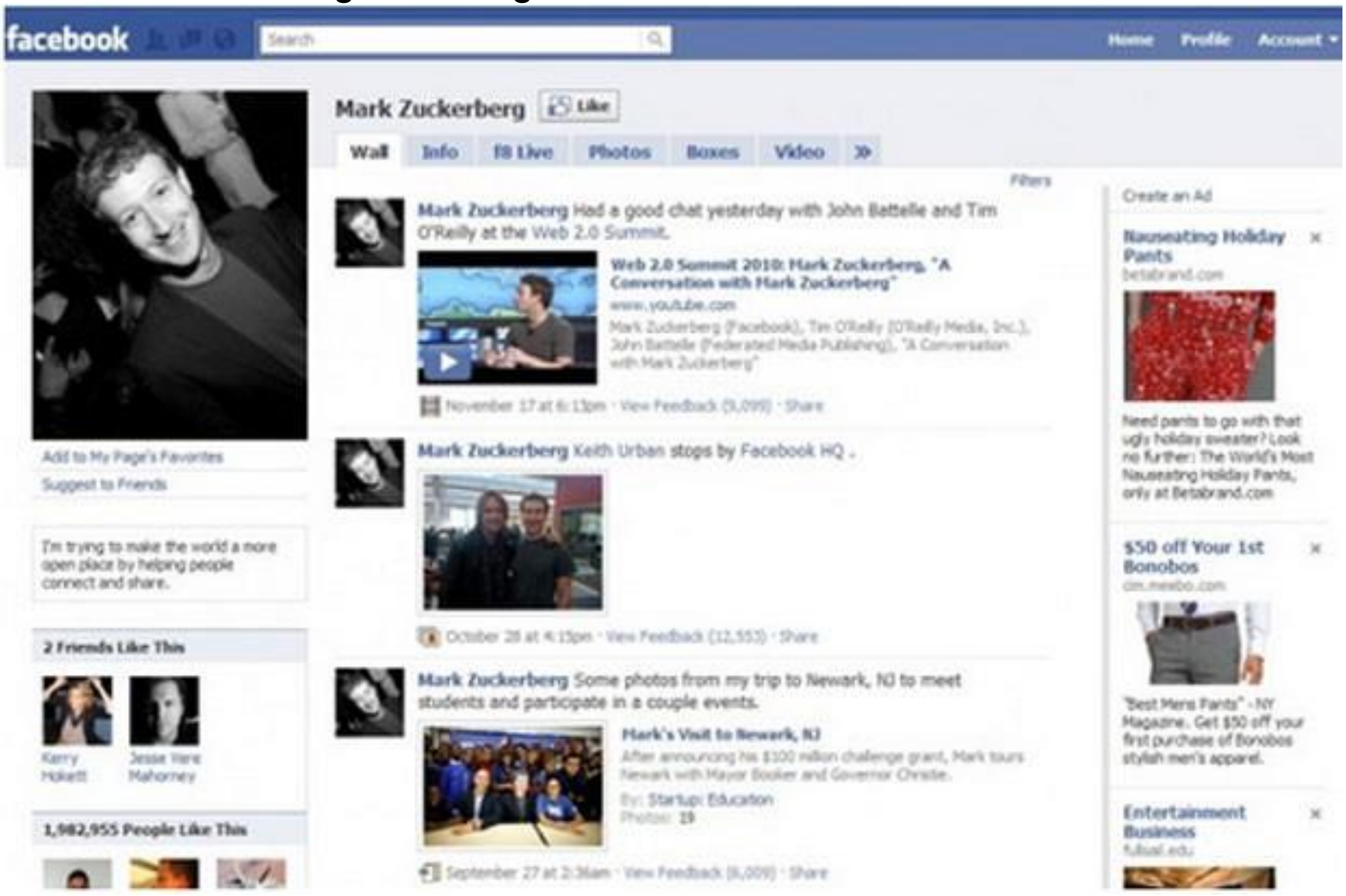

(Fonte: http://s2.glbimg.com/lMsTyUJzjlBd4kmuhkjzan6pVrs=/s.glbimg.com/jo/g1/f/original/2014/02/03/sem-titulo-3.jpg - Reprodução/Wayback Machine)

O *site* também alterou seu *design* das páginas, ganhando um novo nome de “Novo *Facebook*” (*The New Facebook,* em inglês). Uma das novidades eram as abas, onde o usuário poderia alterar sua navegação com facilidade entre notícias, fotos, vídeos, etc.

Também em 2008, a empresa criou a ferramenta de bate-papo e fez o lançamento do aplicativo para os *smartphones*[12] da Apple. Até neste momento só era possível acessar a rede através do navegador do celular, que era mais lento e a visualização não era das melhores comparada a versão para *desktop*. O português do Brasil foi o idioma mais adicionado no *Facebook* em 2008.

O *Facebook* registrou 500 milhões de usuários[13] em julho de 2010, sendo que mais de 20% desses acessavam a rede social por dispositivos móveis (figura 5 mostra a versão para celulares da Apple).

**Figura 5 – Telas da versão móvel do *Facebook* para iOS**

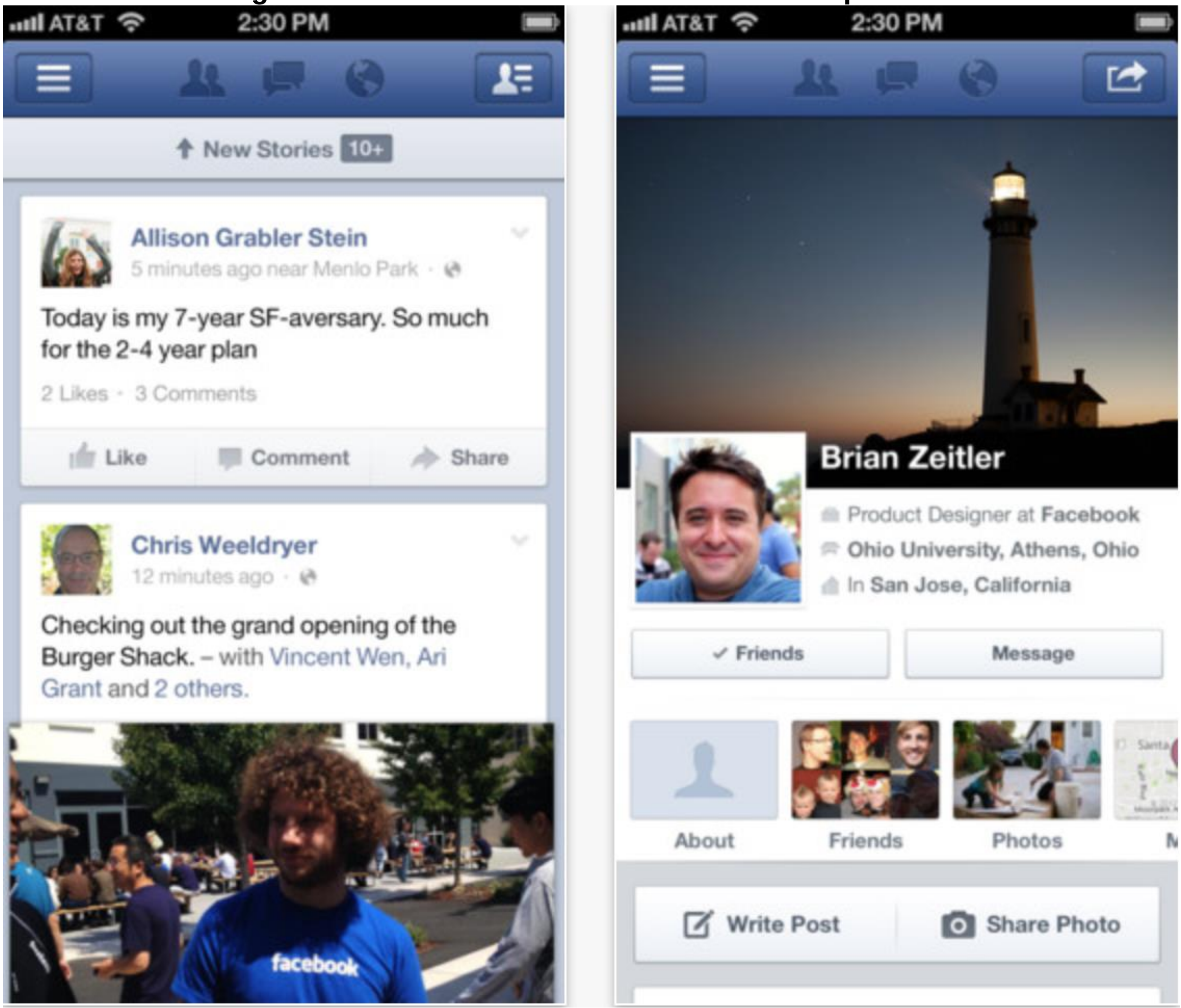

(Fonte: https://9to5mac.files.wordpress.com/2013/03/screen-shot-2013-03-18-at-4-32-46-pm.png)

---

[12] Smartphones são telefones móveis com vários recursos como câmera, acesso à internet, e também a possibilidade de usar aplicativos, pois possui um sistema operacional próprio, exemplo Android ou iOS;
[13] http://g1.globo.com/tecnologia/noticia/2014/02/*Facebook*-completa-10-anos-veja-evolucao-da-rede-social.html

Com a inovação de uma aplicação que fazia o reconhecimento facial em imagens resultou no aumento instantâneo de compartilhamento de imagens e deu a possibilidade de marcar os amigos nas fotos, fazendo com que elas fossem citadas nos murais de atualização.

O ano de 2010 também foi marcado com o lançamento do filme "A rede social"[14] (*Social Network*), que conta a história da criação do *site* e todas as desavenças entre Zuckerberg e os cofundadores da rede.

A aquisição bilionária do aplicativo de compartilhamento de fotos "*Instagram*"[15], foi um dos acertos do *Facebook* em 2012. A implantação obrigatória da linha do tempo do usuário, inclusive para as páginas de empresas, e a inserção de anúncios entre as publicações de amigos e páginas seguidas pelo usuário foram também novidades neste ano. 85% do faturamento de 2011 que o *Facebook* arrecadou, foram de publicidade e venda de anúncios.

Em outubro de 2012, o site alcançou 1 bilhão de usuários ativos. Naquele ano, o jogo que virou febre foi o "*Candy Crush*" (figura 6) e o aplicativo mais comentado foi o "*Tinder*", de encontros.

**Figura 6 – Tela inicial do jogo Candy Crush Saga**

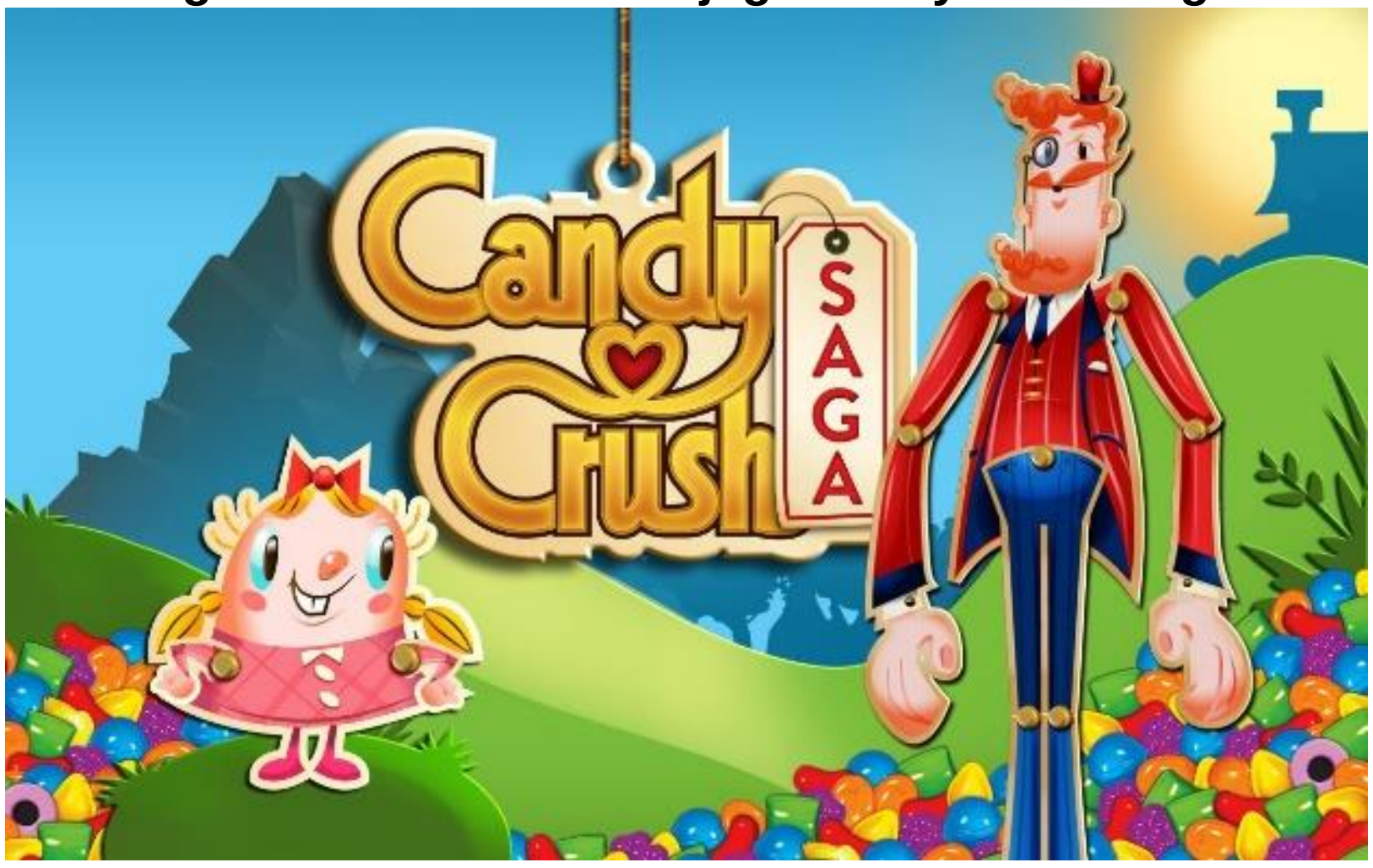

Fonte: http://media.gamerevolution.com/images/galleries/1059/unnamed.jpg

[14] http://www.sonypictures.com.br/Sony/Entertainment/MovieDetails.SonyPictures?id=79214545
[15] http://g1.globo.com/tecnologia/noticia/2012/04/entenda-curta-historia-do-instagram-comprado-pelo-*Facebook*.html

Já em 2013, os acessos móveis ao *Facebook* superaram pela primeira vez as visitas feitas por computador[16]. Isso fez com que Zuckerberg se preocupasse ainda mais com o investimento de aplicativos móveis para a rede social.

Com toda essa trajetória, é nítida a rede com maior expressão mundial com visão e perspectiva de crescimento, uma vez que a intenção da rede é conectar todas as pessoas do mundo. Comprar aplicativos e novas empresas de serviços na internet, criar novas políticas para melhorar o gerenciamento da conta do usuário, dando a ele mais poder de decisão, faz com que o *Facebook* seja uma potência social na Internet.

#### 2.2.1.1 O *Facebook* em números

A seguir, algumas estatísticas impressionantes sobre o *Facebook* no mundo[17]:

- US$ 7,87 bilhões, foi o faturamento da empresa em 2013;
- US$ 1,5 bilhão, foi o lucro líquido da empresa em 2013;
- 1,23 bilhão, foi o número de usuários ativos na rede social em dezembro de 2013;
- 83 milhões de usuários ativos somente no Brasil em 2013;
- US$ 151 bilhões é o valor de mercado do *Facebook*, segundo Nasdaq;
- 4,75 bilhões é o volume de conteúdo publicado por dia no *Facebook*;
- 400 bilhões é o volume de fotos compartilhadas na rede desde 2005;
- 47 escritórios ao redor do mundo;
- 54,2 milhões de páginas na rede social;

No Brasil o site também gera muitos dados interessantes ao olhar de mercado digital, e abaixo segue alguns números sobre o cenário do *Facebook* (dezembro/2011)[18]:

---

[16] http://www1.folha.uol.com.br/tec/2013/08/1326267-brasil-chega-a-76-milhoes-de-usuarios-no-*Facebook*-mais-da-metade-acessa-do-celular.shtml
[17] http://exame.abril.com.br/tecnologia/noticias/os-numeros-do-*Facebook*-dez-anos-apos-sua-criacao#2
[18] http://m.olhardigital.uol.com.br/noticia/toma-essa,-orkut-os-numeros-do-*Facebook*-no-brasil/24134

- A maior faixa etária de usuários está entre 18-24 anos;
- 46% dos usuários são homens;
- A média de amigos por usuário é de 206;
- Conteúdo compartilhado por mês (somente no Brasil):
    - 2 milhões de check-ins
    - 160 milhões de postagens
    - 339 milhões de atualizações de status
    - 460 milhões de uploads de fotos
    - 715 milhões de mensagens
    - 1,6 bilhões de comentários
    - 1,6 bilhões de curtir

### 2.2.2 *Twitter*

Dois anos antes da fundação do *Twitter*, Evan Williams havia sido responsável pela criação do Blogger e, juntamente com Biz Stone, trabalhavam na gigante Google. Saíram de lá para formar a Odeo, uma empresa de *podcasting*[19] que não trouxe resultados em sua área de atuação.

O *Twitter* foi fundado em março de2006 por Jack Dorsey, Evan Williams e Biz Stone como um projeto paralelo da Odeo. A ideia surgiu de Dorsey durante uma reunião de discussão de ideias (*brainstorming*) em que ele falava sobre um serviço de troca de status, como um SMS (Short Message Service).

Chamado simplesmente de Status, o pré-*Twitter* tinha como conceito exatamente o envio de mensagens curtas através do celular, em que você receberia um *twich* (vibração em português) no seu bolso quando um *update* (atualização) era enviado.

---

[19] Podcasting é uma forma de publicação e compartilhamento de arquivos de mídia (áudio, vídeo, fotos, etc) na internet e acompanhar suas atualizações.

Porém, a palavra não ameigou, pois não mostrava precisamente o que era o serviço. Ao buscar nomes parecidos no dicionário, Dorsey e os outros encontraram a palavra *Twitter*, que em inglês tem dois sentidos: “uma pequena explosão de informações inconsequentes” e “pios de pássaros”. Os dois combinavam impecavelmente com o conceito.

O primeiro exemplar do *Twitter* era usado internamente na Odeo, sendo lançado em escala pública e completa (também a versão para computadores) em julho de 2006. Em agosto desse ano, os três fundadores e outros membros da Odeo fundaram a Obvious Corporation, que incluía o domínio “*Twitter.com*”. O microblog se tornou uma corporação separada em abril de 2007, e na figura 7 é apresentada a versão da página neste ano.

**Figura 7 – Página inicial do *Twitter* em 2007**

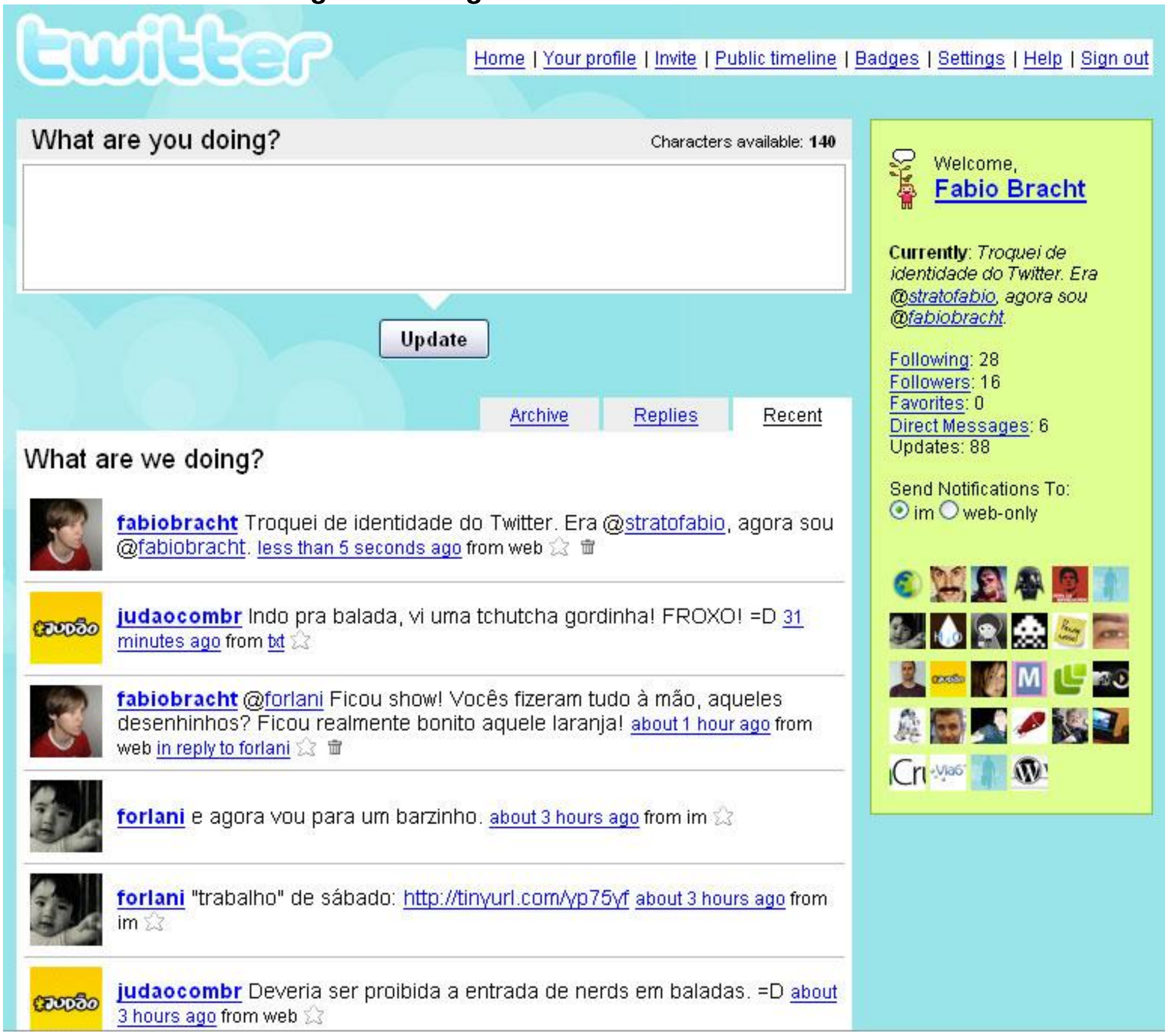

(Fonte: https://h2web.files.wordpress.com/2007/08/*Twitter*_screen.jpg)

A explosão do *Twitter* aconteceu no mesmo ano de 2007 no "*South by Southewest*" (SXSW), um festival de música e filmes para novos talentos, que trouxe a tecnologia como foco através de conferências interativas. O festival chamou a atenção de muitos criadores e empresários da área da tecnologia para apresentar suas ideias.

Neste ano do evento, foram colocadas duas telas de 60 polegadas no principal local de encontro do evento, mostrando exclusivamente mensagens trocadas via *Twitter*. A ideia era que os usuários ficassem sempre de olho no que acontecia durante o evento em tempo real, através da troca de mensagens curtas.

A propaganda e o sucesso durante o festival foi tão grande que o envio de mensagens diárias, que era em média de 20 mil, chegou a 60 mil durante os dias de festa. Assim, os criadores do *Twitter* e a ferramenta receberam o prêmio "Web Award", concedido pelos organizadores do SXSW, ao qual agradeceram em 140 caracteres.

Uma pergunta que paira é do motivo da limitação de apenas 140 caracteres na hora de mandar mensagens pelo *Twitter*. Isso não foi proposital por parte dos criadores. A limitação de caracteres se dá exatamente pelo conceito inicial da ferramenta: uma mensagem SMS.

Além disso, enviar mensagens curtas é o principal foco do serviço e o principal difusor de sites encurtadores de URL[20], como o Bitly[21], Migre.me e outros. A página inicial da ferramenta de encurtadores de links Bit.ly está representado na figura 8.

[20] URL é o endereço de um recurso disponível em uma rede, seja a rede internet ou intranet;
[21] https://bitly.com/

**Figura 8 – Página inicial da ferramenta Bitly**

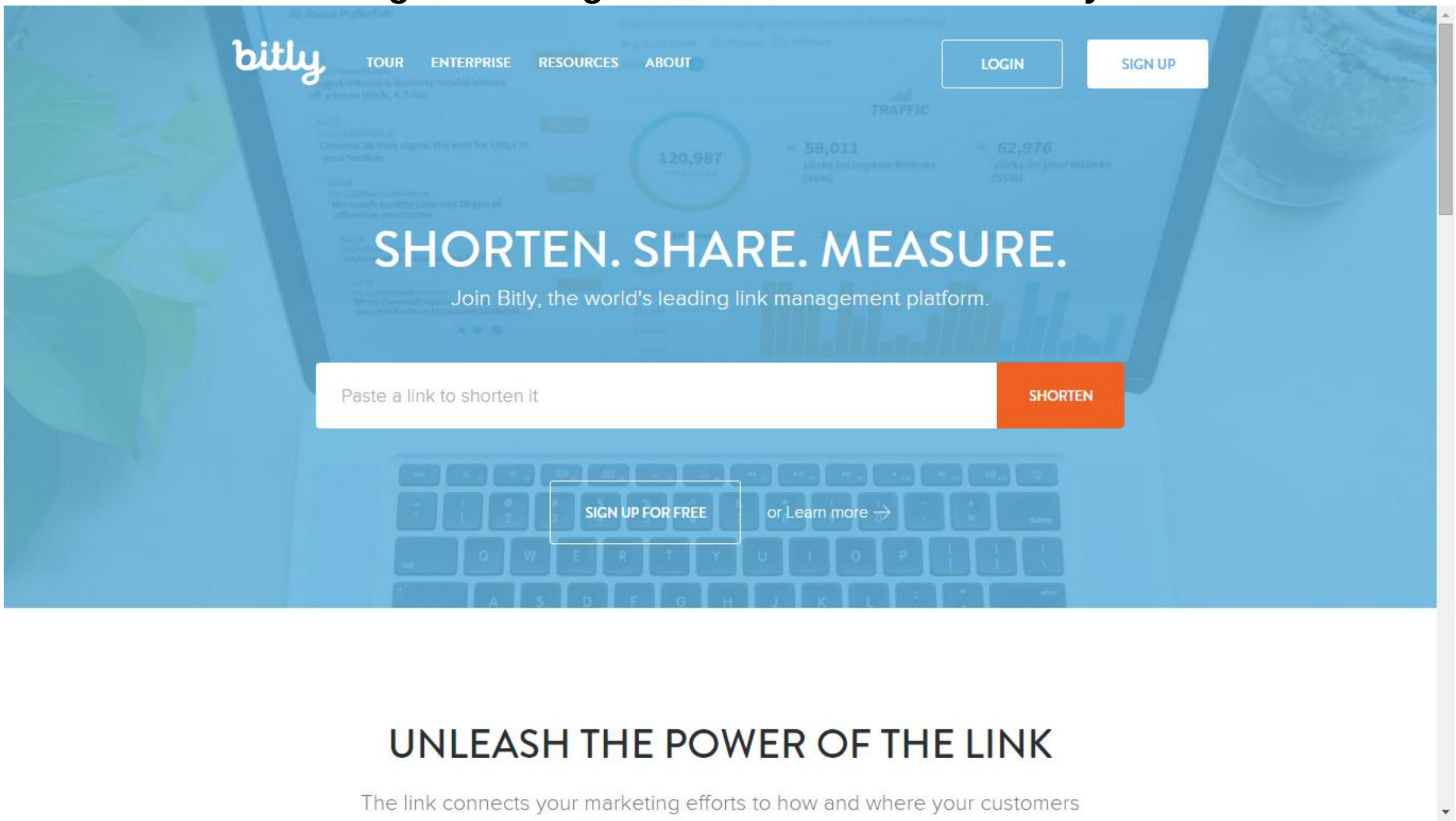

(Fonte: https://bitly.com/)

Um dos principais recursos do *Twitter* não foi lançado logo no começo. Os "*Trending Topics*" (traduzindo, tópicos da moda) trazem os assuntos mais discutidos no mundo do *Twitter* naquele momento. Esse recurso é utilizado também na televisão, onde conseguem "medir" a audiência do programa em tempo real, de acordo com a participação e interação dos telespectadores.

A inserção de uma busca em tempo real de assuntos indexados no sistema se deu em abril de 2009, quando ao observar as marcações feitas pelos próprios usuários, os profissionais do site resolveram incorporar o que antes era um aplicativo em mais uma ferramenta própria através da compra da empresa responsável pelo mecanismo.

Atualmente, o *Twitter* continua sem utilizar propaganda ou anúncios no site, o que pode ser algo um tanto quanto estranho, tendo em vista que nos tempos modernos nada se faz sem a garantia de um retorno.

Entretanto, estima-se que o valor do *Twitter* chegue a um bilhão de dólares, ou seja, o que não é pouco para uma empresa que, teoricamente, não recebe dinheiro por seus serviços. Os mais de um milhão de usuários do microblog certamente agradecem a “falta” de mensagens comerciais, mostrado na versão atual na figura 9.

**Figura 9 – Página inicial do *Twitter* (versão 2016)**

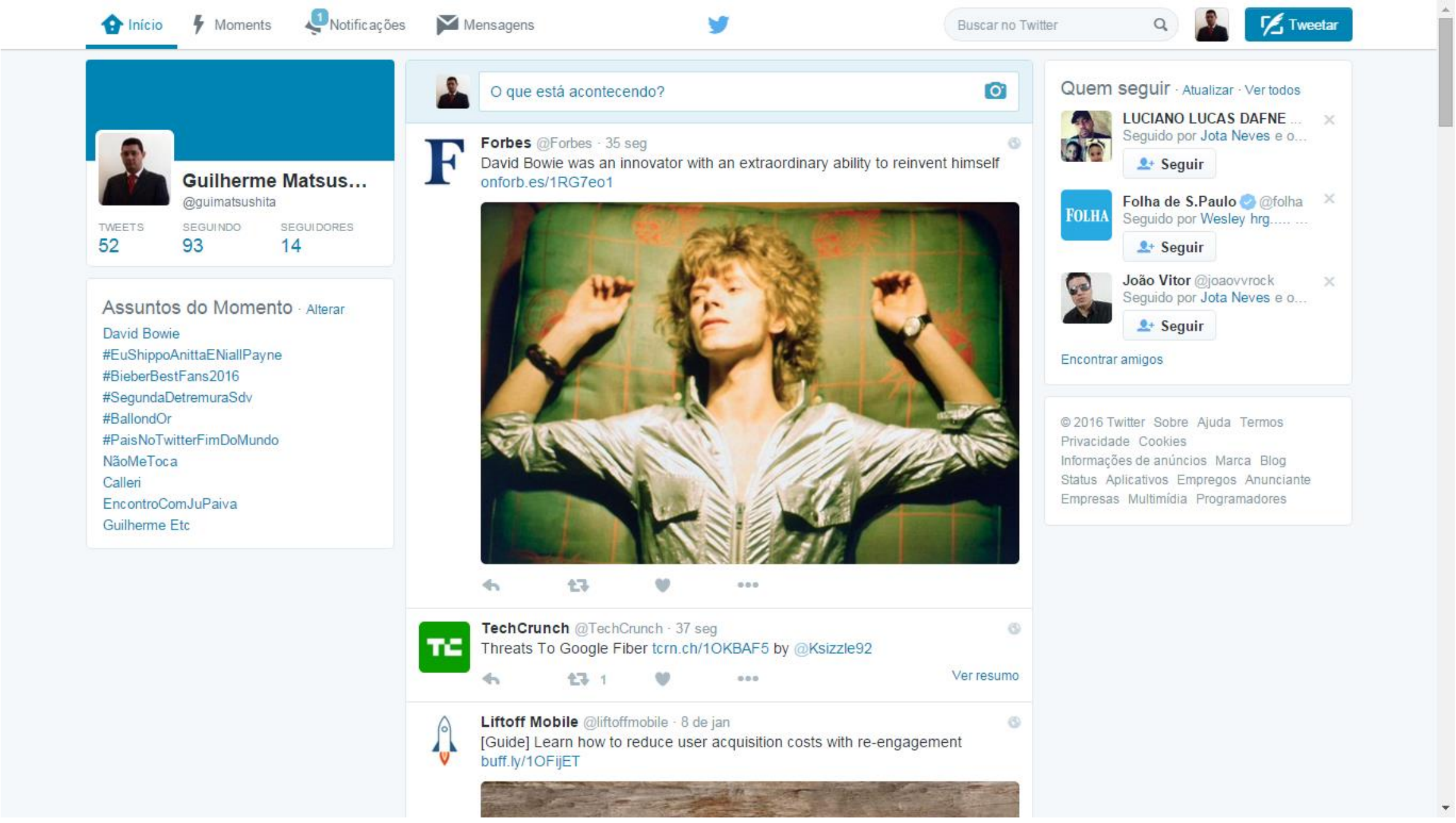

(Fonte: https://*Twitter*.com/guimatsushita)

Além disso, o *Twitter* não ficou parado no tempo. A cada dia surgem novas atualizações e novidades, como as listas de amigos e filtros de “*Trending Topics*” por países. O serviço não parou de evoluir e conta com usuários fiéis, que trocam diariamente cerca de três milhões de mensagens diariamente.

Com crescimento elevado nos primeiros meses de 2009, o *Twitter* parecia estar indo tudo bem, com seus “passarinhos voando para todos os lados”. Entretanto, segundo estudo da HubSpot, a taxa de crescimento havia caído para 3,5% em outubro do ano anterior.

O que se vê também é um crescimento de usuários fora dos Estados Unidos. O português, por exemplo, é a segunda maior língua utilizada no microblog, e o Brasil é o terceiro maior “tuitador”[22] do mundo, atrás de EUA e Inglaterra.

---

[22] Tuitador é o nome informal dado ao usuário ativo da rede social *Twitter*;

Para os criadores da ferramenta, os números mostram que o usuário está utilizando-a com maior consciência e aproveitamento, ou seja, o uso do serviço está mais focado e melhorado. Entretanto, a queda na expansão preocupa, com uma taxa de crescimento muito baixa, de acordo com o blog especializado.

Agora são apresentadas as contas brasileiras com maiores números de seguidores no *Twitter*, atualizado para o ano de 2016, assim mostrada na figura 10.

**Figura 10 – Os 10 usuários brasileiros com maiores números de seguidores no *Twitter***

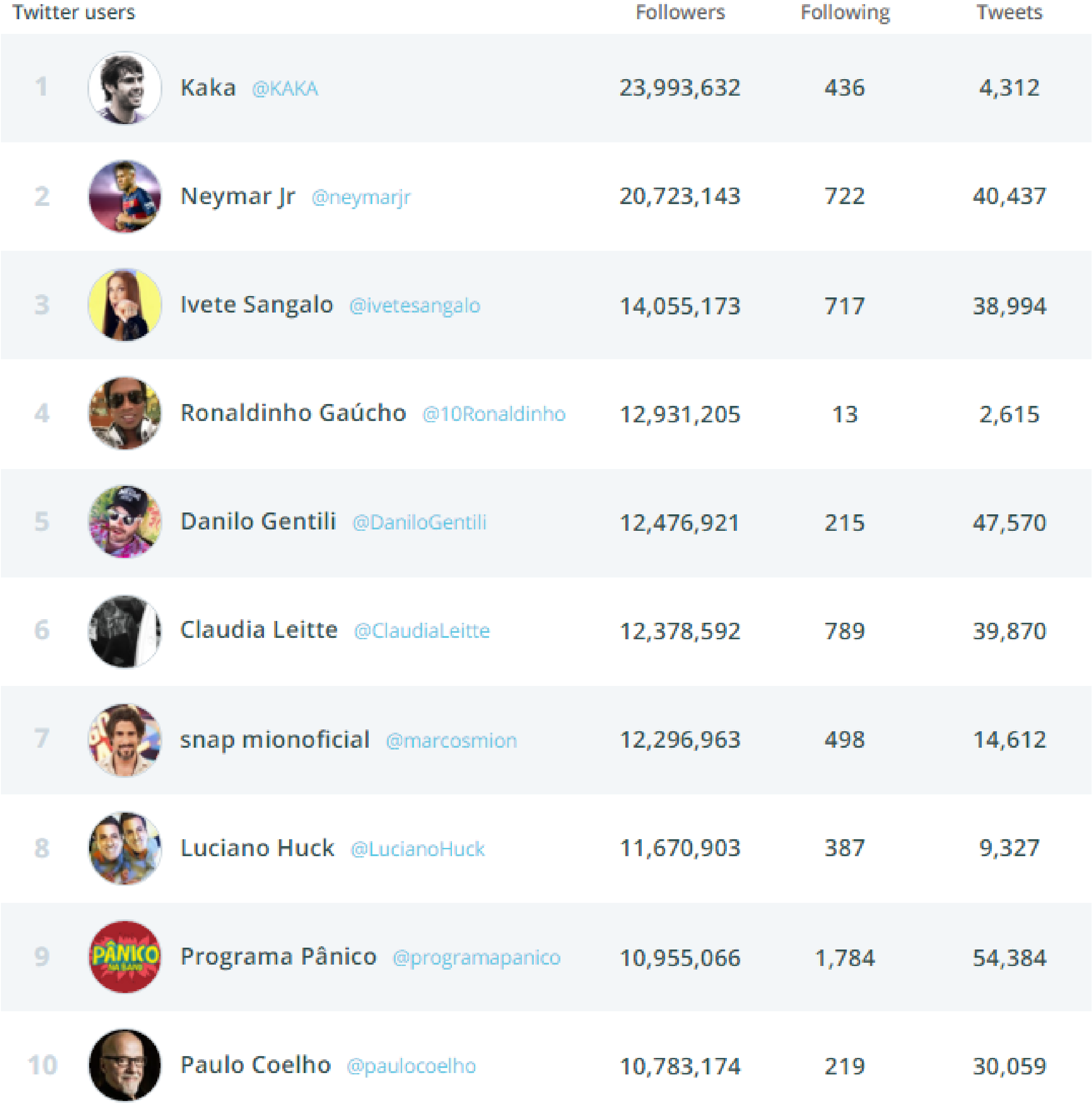

| | Twitter users | Followers | Following | Tweets |
|---|---|---|---|---|
| 1 | Kaka @KAKA | 23,993,632 | 436 | 4,312 |
| 2 | Neymar Jr @neymarjr | 20,723,143 | 722 | 40,437 |
| 3 | Ivete Sangalo @ivetesangalo | 14,055,173 | 717 | 38,994 |
| 4 | Ronaldinho Gaúcho @10Ronaldinho | 12,931,205 | 13 | 2,615 |
| 5 | Danilo Gentili @DaniloGentili | 12,476,921 | 215 | 47,570 |
| 6 | Claudia Leitte @ClaudiaLeitte | 12,378,592 | 789 | 39,870 |
| 7 | snap mionoficial @marcosmion | 12,296,963 | 498 | 14,612 |
| 8 | Luciano Huck @LucianoHuck | 11,670,903 | 387 | 9,327 |
| 9 | Programa Pânico @programapanico | 10,955,066 | 1,784 | 54,384 |
| 10 | Paulo Coelho @paulocoelho | 10,783,174 | 219 | 30,059 |

(Fonte: adaptado de http://twittercounter.com/pages/100/brazil?version=1&utm_expid=102679131-65.MDYnsQdXQwO2AlKoJXVpSQ.1&utm_referrer=http%3A%2F%2F*Twitter*counter.com%2Fpages%2F100%3Fversion%3D1)

#### 2.2.2.1 O *Twitter* em números

- Usuários no mundo: mais de 500 milhões; No Brasil, mais de 46,3 milhões.
- O que é mais postado no *Twitter*:
  - 40,55% de baboseiras e/ou informações inúteis;
  - 37,55% de opiniões e conversas;
  - 5,85% de tuites corporativos;
  - 3,75% de spam;
  - 3,6% de notícias;
  - 8,7% de retuítes

- Somente 20% dos usuários realmente postam tuites;
- Os 5% mais ativos respondem por 3/4 de todo conteúdo;
- Os horários com maior audiência no *Twitter* são, das14h às 17h e das 20h às 22h;
- 69% das empresas tem presença no *Twitter*;
- 55% dos usuários são mulheres;

## 2.3 O volume, a variedade, a velocidade, a veracidade e o valor das informações

### 2.3.1 O volume de dados nas redes sociais

Com as mídias sociais um novo desafio se impõe: hoje as empresas possuem mais dados externos dos seus consumidores do que dados internos. Somos mais de um bilhão de usuários do *Facebook* em todo o mundo. O *Twitter*, com seus 200 milhões de usuários ativos, gera mais de 400 milhões de *tweets*[23] diariamente.

O desafio não é apenas possuir acesso a estes dados mas analisar esta informação de forma diferente. A primeira mudança é: os softwares estão bem mais amigáveis e agora permitem olhar tudo. Ao se deparar para um grande volume de dados é possível utilizar técnicas de estatísticas sofisticadas e identificar padrões de comportamento que não se podia identificar ao olhar apenas para o *buzz*[24] do mês passado.

---

[23] Tweets quer dizer mensagens, formadas por 140 caracteres;

Uma mudança importante é que o mercado passa por uma *datafication*[25], ou seja segundo BERSIN, J. (2014), datificar um fenômeno é colocá-lo num formato quantificável para que ele seja tabulado e analisado. Hoje quanto mais anteparos tecnológicos utilizamos, mais 'datificamos' o nosso dia a dia. Uma simples "bio" do *Twitter* ou um único dia de *checkins* no *Facebook* fornecem muito mais dados hoje do que a humanidade pensou em obter em décadas passadas.

Outra mudança é o fim dos armazéns de dados. Assim como a prefeitura de Nova Iorque precisou colocar em um único armazém de dados as informações que eram pertencentes a diferentes esferas municipais, as empresas também possuem o mesmo desafio. Sem um armazém de dados único que reúna as informações do consumidor espalhadas em todos os softwares utilizados pela empresa será impossível obter *insights*[26] sobre o seu comportamento.

Neste ponto as redes sociais acabam tendo uma grande vantagem competitiva: os dados dos consumidores em relação a várias dimensões do seu comportamento já podem ser encontrados em um só lugar: o que bebem? Aonde vão? O que assistem na TV? O que compram? Tudo isso está nas redes sociais.

A tabela 2 mostra algumas características deste universo de datificar todas as informações obtidas nas mídias sociais.

---

[24] O termo é muito utilizado em marketing, isto é, uma ideia surge e se espalha de maneira diferente através dos comunicadores, contagiando e incentivando as pessoas, surgiu na hora certa, para as pessoas certas no lugar certo;

[25] http://www.hrexecutivecircle.com/pdf/DR14_The_Datafication_of_HR.pdf;

[26] *Insights* quer dizer ideias, neste contexto se trata de uma resposta sobre o comportamento da empresa;

**Tabela 2 – A diferença entre características sobre dados**

| **Já era** | **Like** |
|---|---|
| **Digitalização** | Datafication = datificar um fenômeno é colocá-lo num formato quantificável para que ele seja tabulado e analisado. Datafication é diferente de digitalização. |
| **Amostra de dados** | N = all. Hoje é possível analisar tudo e grande massa de dados. |
| **Dados em silos** | Fim dos silos de dados. Dados de diferentes fontes em um grande armazém de dados. |
| **Dados usados apenas uma vez e depois descartados** | Reuso da informação. O custo de armazenamento em servidores na nuvem permitem guardar um grande número de dados e reusa-los, mesmo que o dado tenha sido recolhido com outro fim. |
| **Why** | *What* é melhor que *why* para *Big Data*. Muitas vezes não vamos conseguir as explicações para tudo mas teremos mais certezas do que falamos graças ao grande volume de dados analisados. |
| **Versão única da verdade** | O fim de uma versão única da verdade. Os dados contam histórias e as histórias dependem das perguntas que se façam. Ou seja, acostume-se a muitas versões da verdade de acordo com as perguntas que você fizer. |

(Fonte: elaborada pelo autor)

O volume dos dados nas redes sociais e os tipos de conteúdo encontrados também é destaque no artigo de Fabricio B. Souza e outros, onde descreve:

> Tanta popularidade está associada a uma funcionalidade comum de todas as redes sociais online que é permitir que usuários criem e compartilhem conteúdo nesses ambientes. Este conteúdo pode variar de simples mensagens de texto comunicando eventos do dia-a-dia até mesmo a conteúdo multimídia, como fotos e vídeos. Como consequência, as estatísticas sobre conteúdo gerado pelos usuários nesses sítios Web são impressionantes. (SOUZA, F.B. et al, 2010)

A figura 11 mostra, de forma ilustrativa, alguns dados sobre o que acontece no ambiente do *Facebook* em um minuto.

**Figura 11 – O *Facebook* em um minuto**

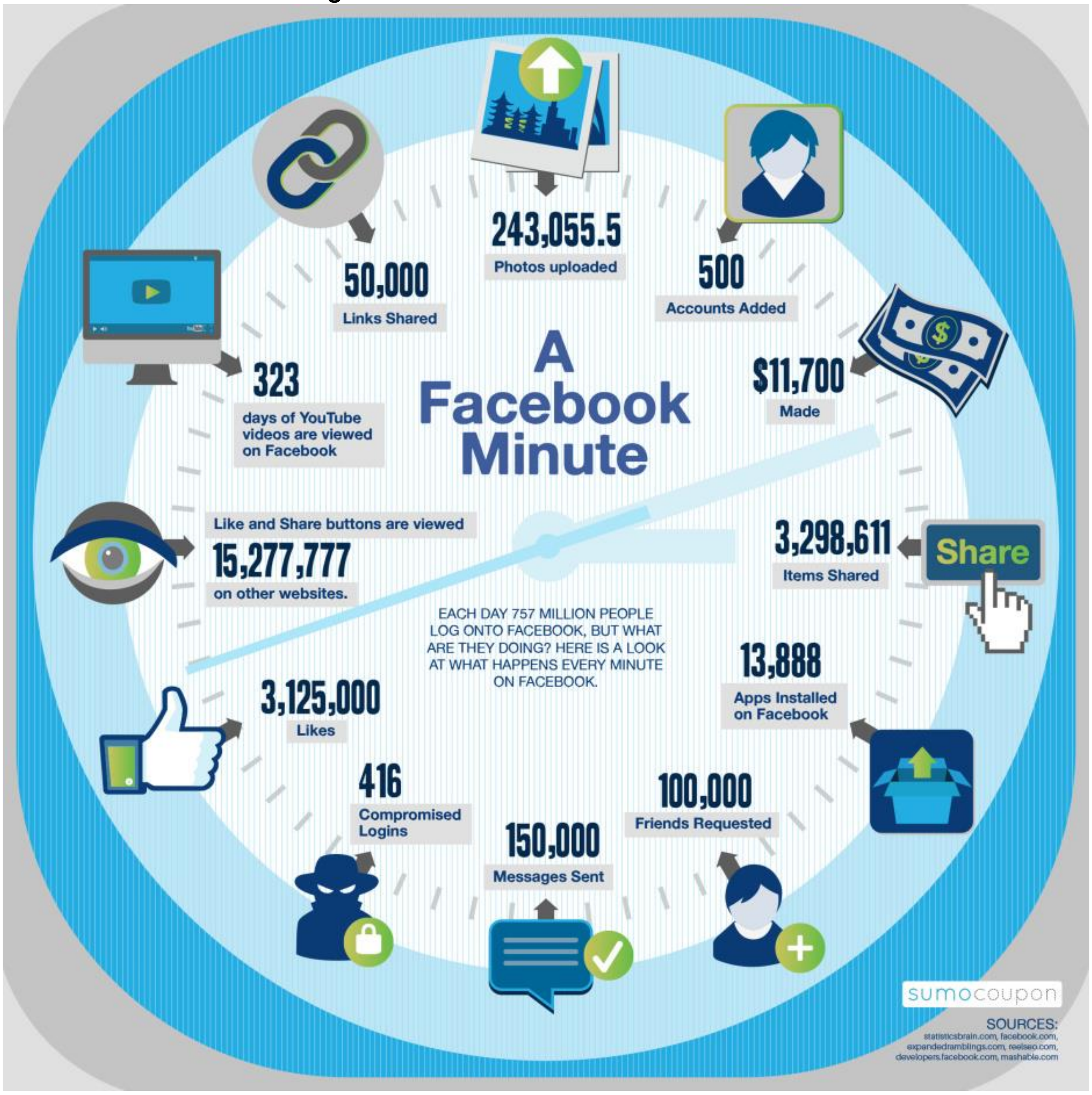

(Fonte: http://zubit.com.br/wp-content/uploads/2014/06/infografia_*Facebook*_1_minuto.jpg)

## 2.3.2 A veracidade das informações nas redes sociais

Nas redes sociais, algumas dúvidas ficam sem respostas a respeito da quão verdadeira é uma informação ou até que ponto tal informação não passa de um buchicho espalhado na rede.

Como saber se uma imagem não é uma fotomontagem ou um clichê ressurgido do céu? Ao capturar qualquer material vindo da internet para cobrir as atualizações do cotidiano em tempo real, os jornalistas, por exemplo, devem verificar a autenticidade da autoria das informações para que não caia em perfis falsos, chamados de *fakes*[27] na rede. Sem o devido cuidado, o erro pode ser catastrófico e pode ser retransmitido imediatamente na velocidade da internet.

Uma pesquisa[28] observada por Naira sobre o uso das redes sociais mostra a questão da confiabilidade das informações e das autorias.

> No caso de busca por fontes, o estudo da Oriella de 2012 concluiu que o uso de mídias sociais na obtenção de novas notícias é uma realidade da maioria dos jornalistas entrevistados, mas somente quando a fonte é conhecida. No Brasil, 52% consultam redes sociais de fontes que conhecem. O uso de blogs é um pouco menor: 44% dos entrevistados 48 disseram ter acessado páginas de conhecidos para buscar novas pautas, e apenas 9% consultam blogs desconhecidos. (CITRANGULO, N. G. p. 48, 2012)

Um dos maiores desafios, para um jornalista, é o de aferir quais das situações correspondem à realidade, por exemplo, se uma postagem no *Twitter* publicado por um estudante qualquer anunciando a prisão de um personagem acusado de assédio sexual em Nova Iorque, ou a publicação da foto de um avião sobre um rio, espalhada por um cidadão americano morador do local.

O uso de tecnologias para a verificação das informações torna este processo mais fácil de trabalhar, segundo Steve Buttry em seu artigo:

> A tecnologia também mudou o jeito de encontrar e lidar com fontes e informações. Conforme participantes e testemunhas de acontecimentos noticiosos compartilham seus relatos pela publicação de textos, fotos e vídeos em redes sociais e blogs, jornalistas podem encontrar e conectar-se mais rapidamente com pessoas que viram as notícias se desenrolarem, tanto usando ferramentas de busca e outras tecnologias, e também pelo *crowdsourcing*. (BUTTRY, S. cap.2, p.15, 2014)

Todos esses procedimentos conhecidos na teoria podem ser úteis, porém na prática as coisas são diferentes quanto tratamos de internet.

---

[27] *Fake* é uma denominação utilizada para denominar contas ou perfis nas redes sociais para ocultar a identidade verdadeira de um usuário;

[28] Disponível em: http://journalismdegree.org/wp-content/uploads/2013/01/Oriella-Digital-Journalism-Study-2012-Final-US-copy.pdf

Um procedimento, para validar uma informação com potencial de notícia da Web, utilizado por vários jornalistas do mundo todo foi elaborado por profissionais de um *site*[29] de conteúdo *online* da França. Segundo os profissionais são cinco etapas a serem seguidas a rigor, são elas:

1. **Identificar o autor do conteúdo**

É a primeira coisa a ser feita. Devemos inicialmente identificar a conta (no *Twitter*, no *YouTube*, no *Facebook*, entre outros) daquele que divulga inicialmente uma informação. Em seguida, verificar se o nome relacionado à conta é um pseudônimo ou o nome verdadeiro dessa pessoa. Por fim, são utilizadas sua conta no *Facebook* ou seu e-mail, para entrar em contato e tentar rapidamente fazer uma entrevista, mesmo via telefone.

Esse primeiro contato permite verificar o elementar, por meio de questões básicas, como:

A. Como a fonte soube da informação publicada? (se uma fonte afirma no *Facebook* que um imóvel no centro de Paris está pegando fogo e ela fala de um telefone do Leste da França, desconfie).
B. A fonte pode dar detalhes sobre o que aconteceu (data, hora, circunstâncias, nome das pessoas envolvidas)?
C. A fonte pode se identificar?
D. A fonte pode detalhar com qual material publicou seu conteúdo on-line?

Julien Pain, jornalista do *site* francês, diz que: “Trata-se da questão da credibilidade que se dá a quem postou um vídeo no *YouTube*. Para ter certeza que foi meu contato quem filmou a cena, eu peço que ele me envie seu arquivo original, antes de ser editado para o *YouTube*. Se foi ele quem produziu o conteúdo, ele deverá ter o arquivo original. Se ele me responde: ‘bem, não sei onde o coloquei’, ele passa a ser uma fonte não confiável.”

---

[29] Disponível em: http://observers.france24.com/fr/

Este mesmo procedimento de checagem é utilizado na BBC, que tem jornalistas atuando numa seção especial: a “UGC Hub” (*User Generated Content*, em português, conteúdos gerados pelos usuários). São esses jornalistas que rastreiam e verificam tudo o que vem da Web, como descreve Nicola Bruno, em sua pesquisa[30] “*Tweet first, verify later*” (Tweet primeiro, verifique depois).

Nesta pesquisa de Bruno também foi analisado por James Morgan, jornalista da BBC que: “o melhor meio para autenticar uma pessoa é falar com ela. Se esta fonte não for legítima, ela evidenciará isso rapidamente ao responder às perguntas factuais”, como por exemplo, “o que é que você vê ao seu redor?.

E, devido às limitações das redes de comunicação, pode demorar horas até que o jornalista consiga falar com a fonte original, sendo assim, o atraso da informação pode ser um risco eminente.

**2. Fornecer o contexto**

Para se compreender um texto, é necessário saber em qual momento ele foi produzido e que situação externa esse texto se refere direta ou indiretamente. A isso chamamos contexto.

Uma foto de um terremoto na China em 2008, publicada por engano para registrar o terremoto do Haiti em 2010. Um vídeo que pretendia demonstrar um massacre na Costa do Marfim em março 2011, quando se tratava de outro estado africano, em data anterior. São casos conhecidos e frequentes que surgem na internet.

Na emissora de televisão francesa “France 24”, este trabalho de contextualização recai sobre uma base de dados formada por mais de 30 mil pessoas espalhadas em todo o mundo, que se inscreveram no *site* da estação

---

[30] Disponível em: https://reutersinstitute.politics.ox.ac.uk/sites/default/files/Tweet%20first%20,%20verify%20later%20How%20real-time%20information%20is%20changing%20the%20coverage%20of%20worldwide%20crisis%20events.pdf

“afirmando que queriam colaborar conosco”, como afirma a jornalista Julien Pain, “das quais três mil já haviam trabalhado conosco”.

E como isso pode ajudar? Se houver um acidente em algum lugar remoto do planeta e seja um acontecimento histórico, a equipe faz o contato com esses colaboradores que mora o mais próximo do local informado e tenta colher informações mais verdadeiras e precisas, em um menor tempo. A jornalista envia o material e o morador diz se as imagens correspondem ao seu ambiente, ou se as roupas das pessoas que aparecem são da região e assim por diante.

### 3. Verificar a informação

A agência francesa AFP determina em suas normas sobre redes sociais que “Devemos sempre verificar as informações factuais obtidas via redes sociais, nos resguardando do risco de toda sorte de histórias inventadas na internet”.

Na agência de notícias norte-americana *Times Union*, próximo a cidade de Nova Iorque, a verificação das informações é baseada “em uma combinação de várias coisas: entrevistar a fonte que relata a história; verificar a história por meio de documentos oficiais, se existirem (documentos jurídicos, boletins de ocorrência policial, estatísticas etc.); e obter detalhes com as instituições, releases ou entrevistas coletivas”.

Claire Wardle, pesquisadora convidada do “*Tow Center”* da Columbia University, diz em seu artigo no “*Verification Handbook*” que devido o fato de qualquer pessoa poder publicar conteúdo variado na internet ligado à algum acontecimento, deixa a imprensa aterrorizada com a possibilidade de publicar boatos falsos e serem enganadas.

> Algumas pessoas procuram enganar propositalmente a imprensa criando websites falsos, inventando contas de Twitter, alterando imagens no Photoshop ou editando vídeos. No entanto, muitas vezes os erros não são deliberados. As pessoas, ao tentar ajudar, às vezes acham conteúdo antigo erroneamente atribuído ao que está sendo noticiado e o compartilham. (WARDLE, C. 2014)

## 4. Decodificar a técnica

Alguns pontos técnicos podem ajudar a evidenciar imagens, vídeos ou conteúdos publicados na internet, de forma a esclarecer e eliminar possíveis erros cometidos por quem os publicam.

Questões como: “A foto foi tirada com qual câmera?”, “A foto foi alterada?”, “A foto foi tirada com *flash*?”. Essas dúvidas são extremamente úteis para compreender a origem de uma imagem e em certas situações são disponíveis por meio de um clique do *mouse* sobre o arquivo original disponível que permite acessar seus dados originais.

Outra opção é fazer o uso de ferramentas disponíveis na internet para validar imagens e outros conteúdos, como disponíveis no manual[31] de verificação feita pelo “The European Journalism” e “Emergency Journalism” em 2014.

A AFP desenvolveu um software chamado Tungstênio, que analisa os dados das imagens para “verificar se um objeto ou pessoa foi removido de uma foto; se um míssil foi duplicado em um foto de guerra; se uma multidão foi adensada; ou se uma imagem foi ‘superdesenvolvida’ para dramatizar”. Este software, bem caro, não resolve tudo, pois os dados podem ser alterados intencionalmente. Entretanto, é um utensílio suplementar na verificação de imagens.

## 5. Monitorar a rede

Monitorar o que se diz na rede, *Twitter*, *Facebook*, blogues e observar o que o estão escrevendo os profissionais e amadores sobre os assuntos recolhidos on-line permitem que o jornalista ou profissional da área de comunicação possa distinguir mais rapidamente aquilo que suscita dúvidas e questionamentos. O exemplo da foto falsa compartilhada de Bin Laden morto fala por si, conforme figura 12.

---

[31] Disponível em: http://verificationhandbook.com/book_br/chapter10.php

**Figura 12 – Montagem de Bin Laden morto**

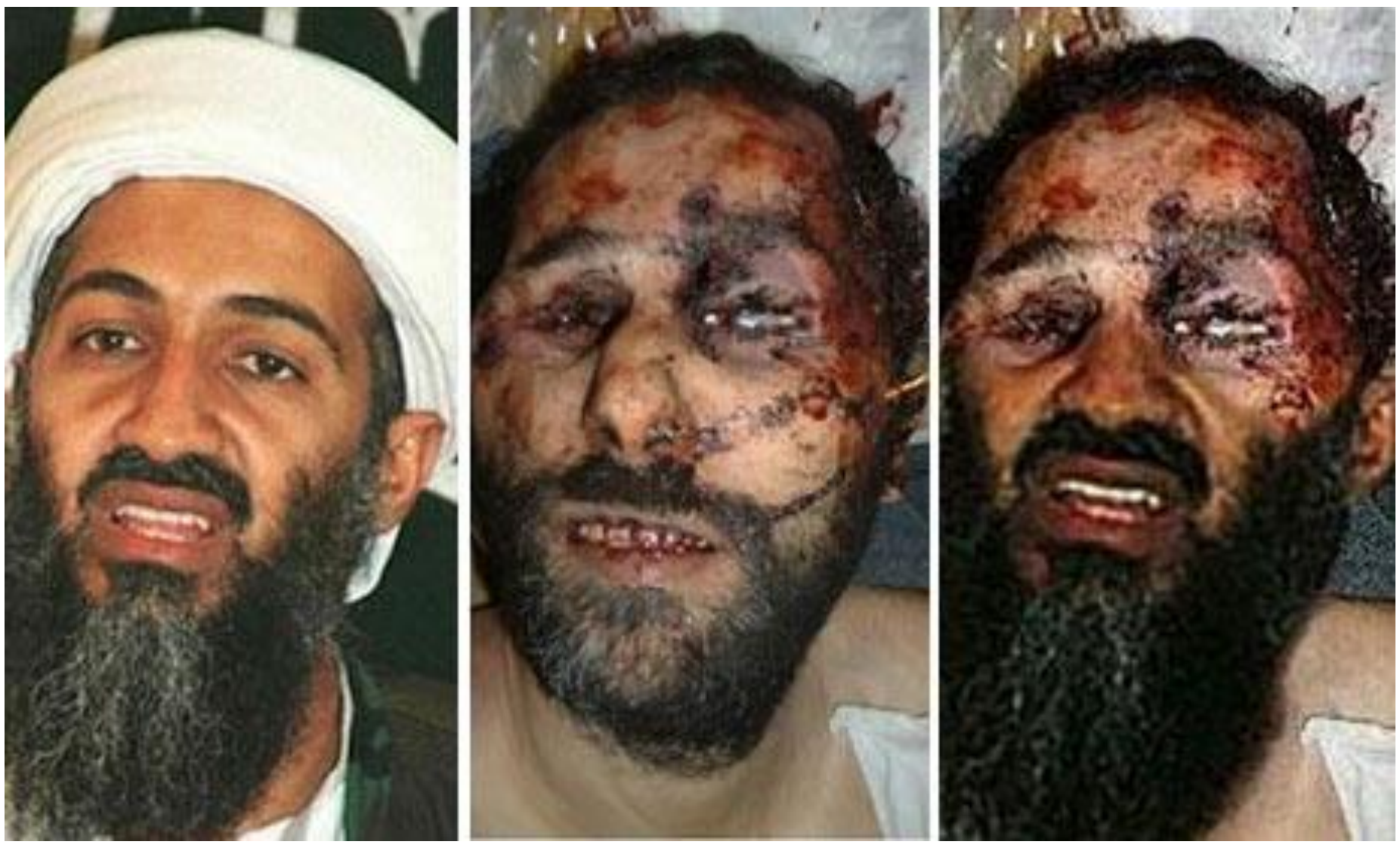

(Fonte: http://exame.abril.com.br/assets/pictures/28788/original_Falso-Bin-Laden-Morto.jpg)

Três conceitos podem ajudar na hora de verificar algum conteúdo na internet, são eles: contexto, conteúdo e código. Os conceitos são a lei dos “3 Cs” definida pelo *blog* britânico “*Online journalism*[32]”. Uma regra que poderá substituir aquela dos “5W”, quem (*who*), o que (*what*), quando (*when*), onde (*where*) e por que (*why*), que os jornalistas utilizam para produzir seus conteúdos.

As redes sociais podem também fornecer informações úteis e valiosas. O problema é que tudo na internet ocorre em uma velocidade que não é possível acompanhar e não se consegue diferenciar o que é verdade ou mentira com a mesma velocidade. Isso torna mais difícil reagir a rumores, por exemplo, impedindo que serviços de emergência invalidem uma mentira para manter a tranquilidade em uma dada situação.

### 2.3.3 O valor das informações

O valor das informações das redes sociais como *Facebook*, *Twitter* ou *YouTube*, tem um peso fundamental para as estratégias da corporação na forma como elas interagem no ambiente empresarial, descritos no texto da dissertação de mestrado de Flavia Ivar de Souza (2012):

> Essas ferramentas foram incorporadas às empresas de forma progressiva. Esse fato facilitou o trabalho em conjunto, o compartilhamento de informações, o acesso a aplicativos previamente existentes e a disseminação de conteúdo. O uso inteligente dessas tecnologias sociais permite que haja mudanças nas formas como as organizações interagem com seus *stakeholders* (clientes, fornecedores, acionistas, funcionários, cidadãos, governo e sociedade em geral).

---

[32] Disponível em: http://onlinejournalismblog.com/

A informação é um recurso de grande valor para a organização comparada a qualquer outro recurso material, financeiro ou de produção e está diretamente vinculada ao mercado e a geração de lucros, na visão de Danielly e Sirlene, citando MORESI (2000):

> A informação é um recurso valioso que faz o diferencial na competitividade de qualquer organização. Ela pode ser comparada, em termos de importância, a qualquer outro recurso da organização, seja material, de produção ou financeiro, e está diretamente ligada ao destaque no mercado e à lucratividade de uma instituição. (INOMATA, D.O., Revista ACB, v.18, n.2, p.1004, jul./dez., 2013)

Assim, se dá por encerrado este capítulo tratando sobre as redes sociais, focando o *Facebook* e o *Twitter*. Segue então o próximo capítulo que fala mais sobre o monitoramento com foco no retorno estratégico empresarial.

# Capítulo III

## O monitoramento e o retorno estratégico empresarial

# 3 O MONITORAMENTO E O RETORNO ESTRATÉGICO EMPRESARIAL

## 3.1 Introdução ao monitoramento das redes sociais

A proliferação dos dados vem causando grande impacto em nossas vidas. Computadores e dispositivos móveis, como telefones celulares, tablets permeiam em nossas atividades diárias uma grande quantidade de dados sobre nossos clientes, fornecedores e sobre suas operações negociais. A quantidade de soluções e aplicativos, principalmente para os dispositivos móveis (conforme figura 13), está cada vez mais presente nas relações sociais do mundo moderno.

**Figura 13 – Exemplos de aplicativos para redes sociais do Google Play**

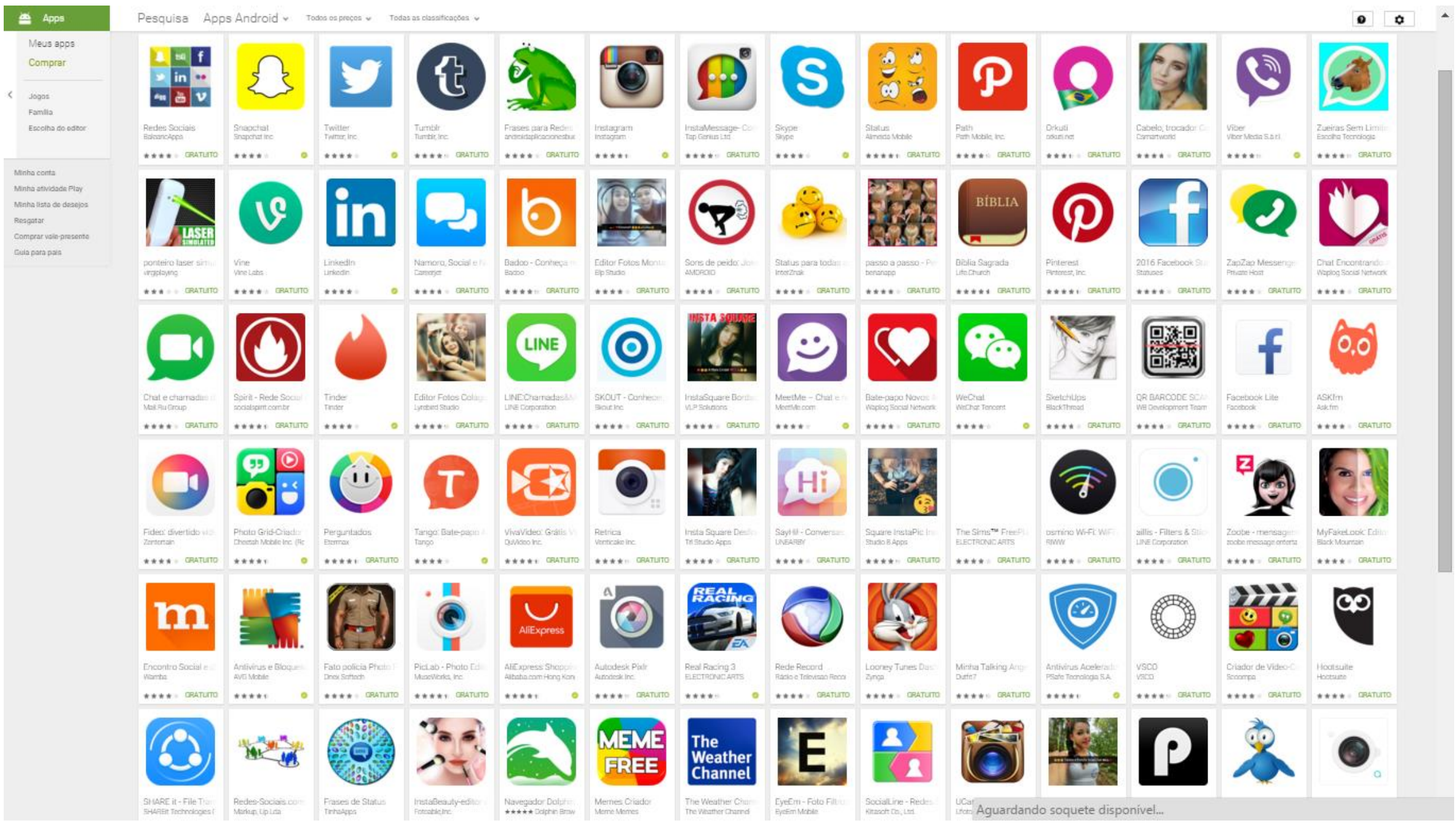

(Fonte: https://play.google.com/store/search?q=redes%20sociais&c=apps&hl=pt-BR)

Para RAMOS & BRASCHER (2009), a velocidade e a amplitude com que as informações são geradas provocaram a necessidade de técnicas para tratamento de grandes massas de dados e informações. Devido à importância da informação para geração de conhecimento e consequente identificação de oportunidades para inovação, a presente seção tem como objetivo demonstrar a utilização de ferramentas de visualização de informação. Objetiva-se extrair informação relevante da massa de informação e desta forma auxiliar na descoberta de oportunidades nas redes sociais e em bases de dados.

### 3.2 O processo de monitoramento

As etapas para um monitoramento das redes sociais de forma eficaz, segundo Salustiano, são:

- **Planejamento** – Identificar onde, como, quais redes sociais serão monitoradas, o período para conseguir os dados consistentes e o alcance do público a ser analisado.
- **Busca do conteúdo** – Buscar os conteúdos nas redes pré-escolhidas segundo a relação com o produto e/ou serviço a ser monitorado.
- **Análise das informações** – Identificar quais conteúdos são relevantes para a empresa.
- **Classificação** – Identificar se o conteúdo é positivo, negativo ou neutro; Se foi espontâneo ou estimulado; Se foi próprio ou replicado; Se foi originado por problema pré-existente ou de forma avulsa.
- **Consolidação** – Estruturar os dados obtidos segundo sua classificação, transformar os resultados em gráficos para que possam ser interpretados.
- **Interpretação** – Segundo o *briefing*[33], analisar os dados para responder as perguntas iniciais do cliente.
- **Análise** – Estruturar todos os dados no formato de uma análise precisa sobre a percepção do usuário em relação à marca, além de sugerir formas de abordagem e ações de marketing, por exemplo.
- **Relatório** – Desenvolver o relatório final para a entrega ao cliente.

---

[33] O Briefing é um complexo de informações (indicações do que o cliente pretende) e também uma passagem de dados que ocorre geralmente durante reuniões, a fim de desenvolver um trabalho que descreve a situação de uma marca ou empresa (seus problemas, oportunidades, objetivos e recursos). IKEDA, A.A, BACELLAR, F.C.T. 2004.

Segundo NARESSI, L. (2013) alguns pontos devem fazer parte do processo de monitoramento das mídias sociais de forma genérica, conforme demonstrado na figura 14, desde a identificação das conversas até a geração de conhecimento.

**Figura 14 – Processos envolvidos no monitoramento de redes sociais**

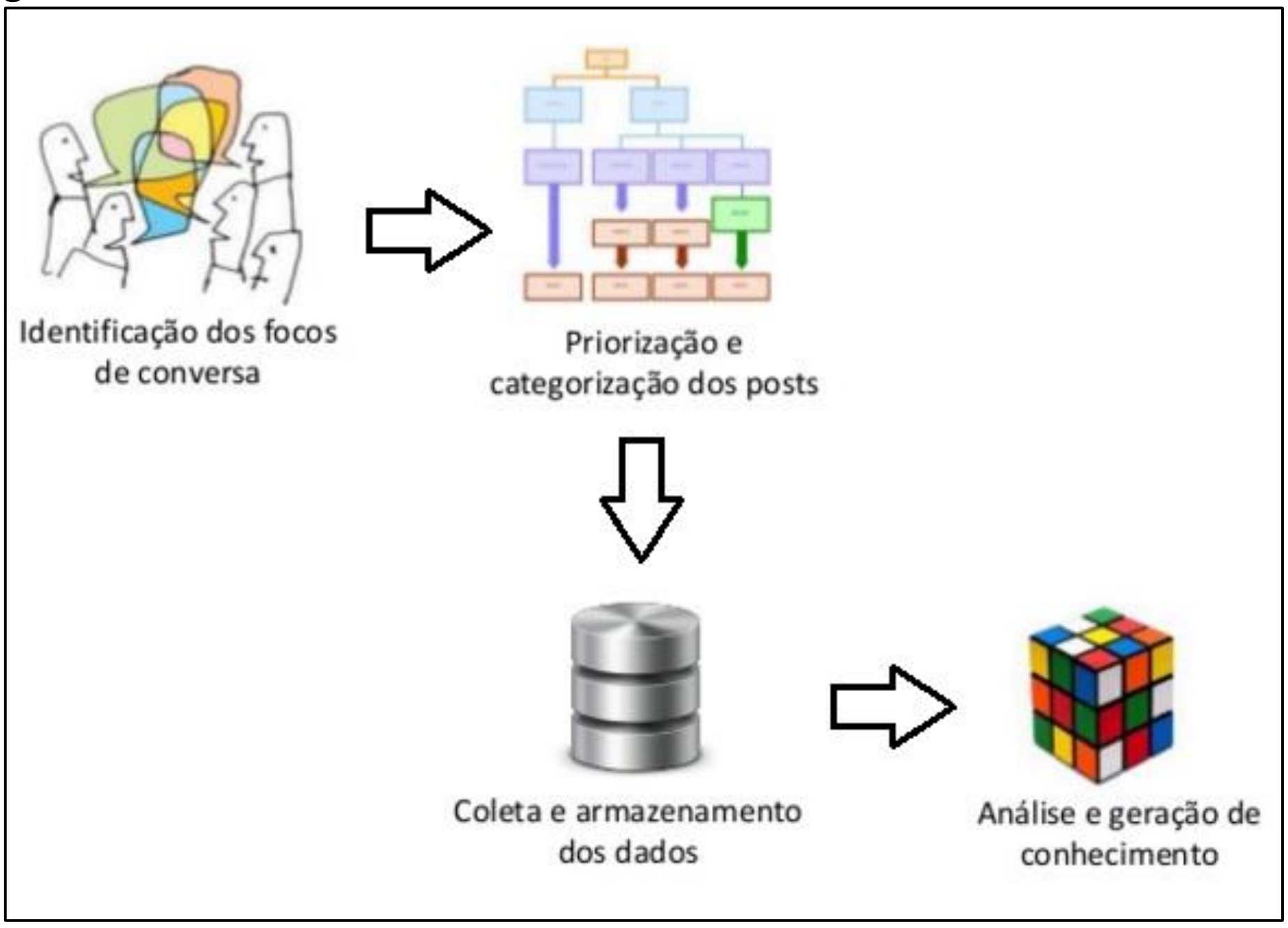

Fonte: adaptado de NARESSI, L. (2013) (http://pt.slideshare.net/leonaressi/metricas-redes-sociais-e-buzz-intelligence)

O monitoramento busca entender o que o consumidor fala, como fala, porque fala e para quem fala.

No cenário de desenvolvimento Web, uma API [34] é caracteristicamente um conjunto de tipos de requisições HTTP[35] junto com suas respectivas definições de resposta. Em aplicações de redes sociais online é corriqueiro encontrarmos APIs que catalogam os amigos de um usuário qualquer, suas comunidades que participa, entre outras informações relevantes. São recursos complementares que trabalham em paralelo com as redes sociais em função de uma determinada tarefa. *Facebook* e *Twitter*, por exemplo, são as redes que utilizam esses recursos extras para a obtenção de dados dos usuários.

[34] API provém do Inglês *Application Programming Interface*, que significa Interface de Programação de Aplicações;

[35] HTTP significa *Hypertext Transfer Protocol*, em português Protocolo de Transferência de Hipertexto;

Segundo SOUZA (2010) em seu artigo, as APIs são perfeitas para a coleta de dados de redes sociais online, pois oferecem os dados em formatos estruturados como XML e JSON. Além dessa função, o *Twitter*, por exemplo, oferece várias funções em sua API. Com o uso desse complemento é possível coletar 5000 seguidores de um usuário através de uma única requisição. A coleta dessa informação através do sítio Web convencional necessitaria de centenas de requisições, visto que o *Twitter* só mostra alguns seguidores por página. Além disso, cada página teria uma quantidade muito grande de dados desnecessários, que deveriam ser tratados e excluídos.

O autor ainda destaca que vários sistemas possuem APIs, incluindo *Twitter*, Flickr, *YouTube*, Google Mapas, Yahoo Mapas, etc. Com tantas APIs existentes, é comum ver aplicações que utilizam duas ou mais APIs para criar um novo serviço, que é o que chamamos de *Mashup*[36]. Uma interessante aplicação chamada "Yahoo! Pipes[37]", permite a combinação de diferentes APIs de vários sistemas para a criação automatizada de *Mashups*.

O ciclo de vida da mídia social está atrelado ao processo de monitoramento da rede social, uma vez que o ciclo tem etapas que completam o monitoramento no seu entendimento conforme apresentado na figura 15:

[36] Aplicação Web que utiliza conteúdo de mais de uma fonte para criar um novo serviço completo.
[37] http://pipes.yahoo.com/pipes

**Figura 15 – Ciclo de vida da mídia social**

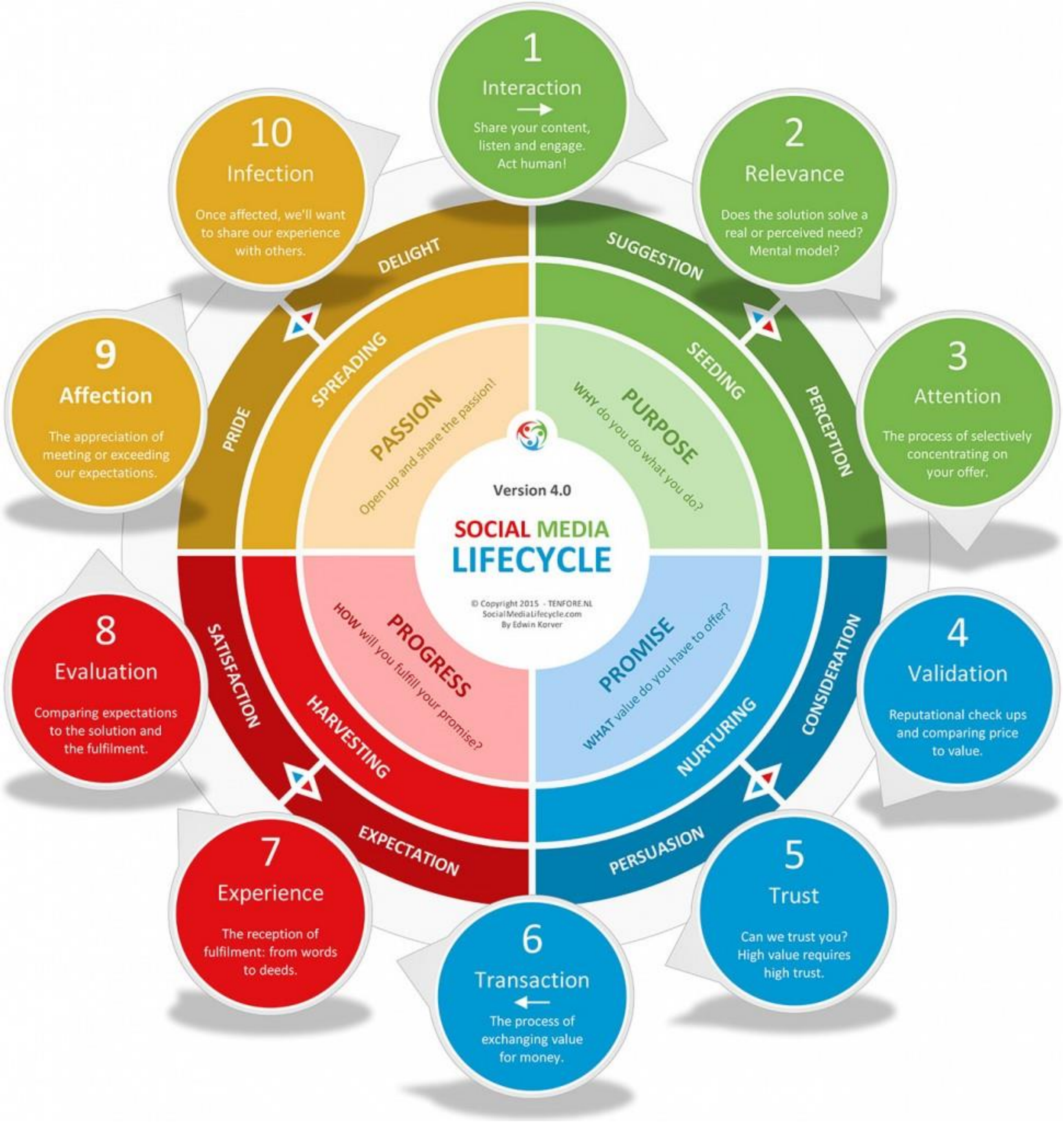

(Fonte: http://tenfore.nl/wp-content/uploads/2014/09/socialmedialifecycle-v40-13001-970x1024.jpg)

## 3.3 As métricas das redes sociais

Nos dias atuais o ambiente de marketing digital traz inúmeras formas de mensuração, com acesso a uma vasta quantidade de informações e conhecimento, entretanto é preciso ser cuidadoso ao escolher quais as formas de medir que vão ser utilizadas nos processos das redes sociais, pois pode haver riscos de ficar abarrotados de informações que não irão fazer sentido para a empresa.

É preciso saber sempre em favor do que é trabalhado e alinhado e, novamente, o objetivo principal é essencial nesse momento. Ao analisar o universo das mídias sociais, mais especificamente as redes sociais, as principais métricas que podem ser utilizadas são, segundo artigo de MOURA; OLIVEIRA (2015):

- Métricas relacionadas aos canais próprios da marca (páginas, perfis, blogs);
- Métricas relacionadas à reputação digital;
- Métricas relacionadas à conversão e ROI;
- Métricas relacionadas às ações da marca no ambiente digital, como campanhas, promoções e aplicativos.

Segundo o "Comitê de Social Media do IAB Brasil[38]", em um esforço para padronizar os tipos de métricas utilizadas em redes sociais, podem ser destacadas as cinco principais dimensões de métricas: base de relacionamento, alcance, engajamento, atendimento e transações, explicado na tabela 3.

**Tabela 3 – Tipos de métricas utilizadas nas mídias sociais**

| Dimensão | Descrição das métricas | Exemplo |
|---|---|---|
| **Base de relacionamento** | Relacionadas às quantidades de usuários que seguem uma página, canal ou perfil em rede social. | Quantidade de fãs no *Facebook*, seguidores no *Twitter*, inscritos no *YouTube* |
| **Alcance** | Relacionadas às quantidade de pessoas que foram impactadas por seus conteúdos e ações nas mídias sociais. | Alcance orgânico, alcance pago, alcance viral, views |
| **Engajamento** | Relacionadas ao envolvimento do público com o seu conteúdo, ação ou página. | "Curtir", comentários, compartilhamentos, cliques, menções à *hashtag*, etc |
| **Atendimento** | Relacionadas ao serviço de atendimento ao consumidor via redes sociais (SAC 2.0), pois uma marca presente neste ambiente deve estar aberta ao diálogo. | Taxa de resposta, tempo de resposta, quantidade de menções que demandam atendimento, índice de solução, reversão de sentimento dos casos de SAC, etc. |
| **Transações** | Relacionadas as atividades realizadas por meio de recursos que não são nativos das redes sociais | Vendas, participação em concursos, downloads, geração de *leads,* etc. |

(Fonte: elaborada pelo autor)

[38] http://iabbrasil.net/comite/social-media

Percebe-se pelos exemplos na tabela 3 que as métricas, são dados brutos que podem ser coletados em ferramentas como *"Facebook Insights"*, *"Google Analytics"*, o próprio SCUP ou outros softwares de monitoramento e mensuração disponíveis no mercado.

Os KPIs, por sua vez, são indicadores definidos pelos gestores para acompanhar o desempenho das métricas associadas ao objetivo do negócio como um todo. O KPI (*Key Performance Indicators*, em português, Indicadores de Performance[39]) está associado ao negócio e aos objetivos do cliente, oferecendo contexto à métrica e trazendo consigo uma meta que visa, em última instância, alcançar o objetivo definido lá no início do processo.

O KPI é uma métrica ou conjunto de métricas que está diretamente relacionada com o sucesso da ação que está sendo mensurada, e deve estar associado a uma tomada de decisão. (MOURA, C., OLIVEIRA, M. 2014) Desta forma, o KPI é repetidamente composto por índices, comparativos, faixas de valores, podendo ser avaliado a partir das métricas fornecidas inicialmente.

É de extrema importância deixar explicado que o KPI deve estar relacionado à referências, tanto da marca própria, como de um concorrente ou segmento de atuação no mercado. Imagine a seguinte situação: e a média de menções negativas de uma determinada ocorrência for de 25% do total e a empresa atingiu somente 10% deste universo, mostra que o KPI determinado anteriormente indica um nível satisfatório em relação à concorrência e que as estratégias de CRM ou marketing nas redes sociais estão retornando bons resultados.

Continuando no raciocínio de Moura (p.11, 2014), de forma geral, enquanto as métricas trazem dados inteiros e separados, os KPIs dão um melhor conhecimento sobre o cenário das métricas, sendo:

---

[39] Performance neste caso se refere à melhoria de desempenho ou variação de resultados obtidos em um determinado processo;

- Determinados a partir dos objetivos;
- Calculados através das métricas;
- Associados a uma meta;
- Dirigidos a uma tomada de decisão;

Mensurar e avaliar os resultados encontrados. Quem é gestor, pesquisador ou outro profissional preocupado sabe da importância desta máxima como forma de delimitar as ações, auxiliar nas tomadas de decisões e entender melhor e com eficiência todos os processos envolvidos, segundo Felipe Raffani[40].

O que preocupa os profissionais de mídias sociais digitais é o fato de que as métricas passaram a ser mais importantes do que as próprias ações em si. Marcas preocupadas por um número absurdo de "*likes*" no *Facebook* ou seguidores no *Twitter*, como se esses números representassem sucesso perante os consumidores. Entretanto essas empresas deviam se preocupar com a interação e os resultados oriundos das ações.[41]

Determinadas dúvidas ainda devem ser respondidas sobre o universo das redes sociais e suas estatísticas. Sua empresa tem fãs reais? Métricas são apenas números sobre um determinado assunto. Nos dias de hoje, vimos muitas pessoas sugerindo a sua rede de amigos virtuais no *Facebook* para que "curtam" certa página de empresa ou produto, e os amigos clicando no "curtir" somente pela amizade e não pela preferência ou opinião própria sobre a página, ainda mais com a facilidade da mobilidade (conforme figura 16). Ou seja, não existe uma relação real do usuário com a empresa, não há nenhum engajamento. Um número que não diz nada para a análise estratégica da organização.

[40] http://iabbrasil.net/artigo/para-metricas-fazerem-sentido
[41] [1] http://pt.slideshare.net/natraldi/mtricas-e-monitoramento-no-*Facebook*

**Figura 16 – Usuário acessando o Facebook no celular**

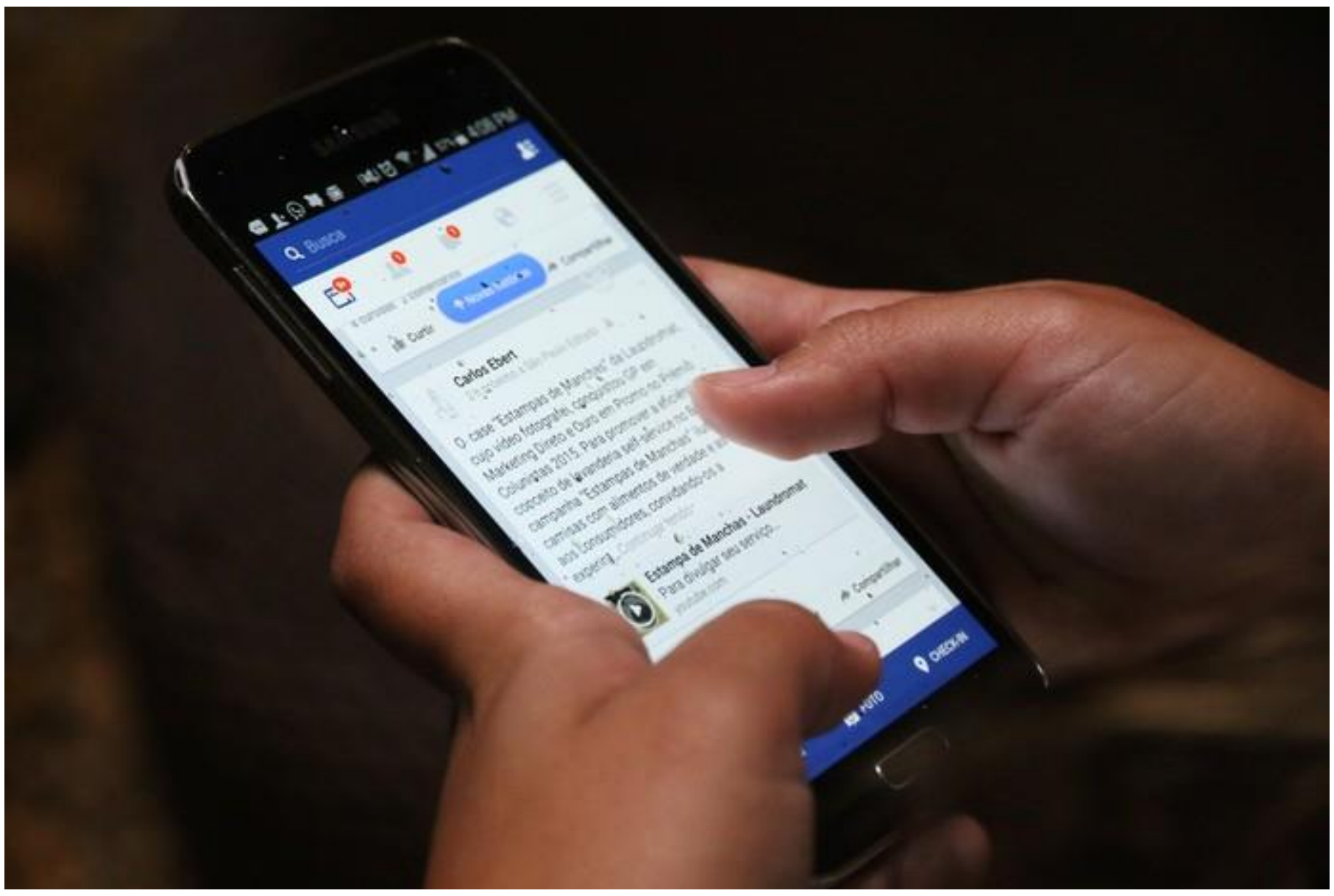

(Fonte: Luciana Maline/TechTudo - http://www.techtudo.com.br/dicas-e-tutoriais/noticia/2016/01/como-fazer-transmissoes-ao-vivo-no-facebook-usando-o-celular.html)

Um estudo recente conduzido pela consultoria Booz&Co e a Buddy Media a respeito do impacto do "*social media marketing*" apontou que a maior parte das pequenas e médias empresas ainda não sabe usar plenamente essas ferramentas.

Um fato é de que as organizações precisam se empenhar para melhorar três competências: desenvolvimento de conteúdo, gerenciamento de comunidades e análise de métricas, ou a interpretação em tempo real dos números e resultados.

Ter o domínio dessas competências significa poder ter uma grande oportunidade para a construção de relações poderosas e duradouras com os clientes, aproveitando a grande popularidade vinda das plataformas digitais como *Facebook*, *Twitter* e *YouTube*.

A análise dos números se dá em quatro níveis de expertise — de uma simples "contagem" da atividade nas mídias sociais até alcançar objetivos estratégicos, segundo a pesquisa da Booz&Co.:

1. Alcance – o alcance trata de identificar quantos fãs, seguidores, assinantes ou visitantes sua empresa/marca possui. Também é preciso saber quantos *posts*[42] foram comentados ou compartilhados, em quais *sites* sua empresa está sendo comentada e por fim, quantas menções estão sendo feitas em diferentes canais na Web;

2. Engajamento – esta fase de engajamento ultrapassa a contagem de fãs apenas. A empresa já participa de comunidades e fóruns com ligação com a marca e o tipo de discussão que acontecem nesses grupos. Também se inicia o trabalho de analisar padrões nas mensagens, comentários, compartilhamentos, entendendo o que orienta a participação e a distribuição dos conteúdos.
3. "Vestir a camisa" – neste momento do processo é feita a identificação e estimulação dos comportamentos dos usuários que podem ser considerados como um verdadeiro comprometimento com a marca da organização. Esta etapa inclui métricas como a predisposição para realizar recomendações, compartilhar conteúdos e fazer comentários a favor da marca ou "defender" a marca, ou seja, fazer elogios ou advogar em favor da empresa.

4. Retorno sobre investimento – esta é a última etapa e também considerada a mais complicada e por consequência, a mais difícil de ser implementada. Devem-se criar indicadores de desempenho (KPI – *Key Performance Indicators*) que garantem a efetividade do investimento feito nas mídias sociais. Este processo é complexo, pois existem pontos como propaganda boca a boca que são difíceis de serem mensurados. Por outro lado é possível verificar a taxa de conversão que mostra quantas pessoas que navegam no site e realizam algum tipo de ação, e também o incremento em vendas vindas das redes sociais.

O mais importante é pensar que conteúdos interessantes e eficientes ainda são a melhor maneira de incentivar as pessoas a compartilharem e se tornarem fãs reais da sua marca.

[42] Um *post* é uma mensagem, notícia ou texto publicado em um *blog*, *site* ou página de relacionamento como o *Facebook* ou *Twitter*;

### 3.4 A análise das informações obtidas

De acordo com MARTELETO (p.71-81, 2001), a análise das mídias sociais é realizada do geral para as partes, ou seja, da estrutura da rede para a relação do indivíduo, do comportamento para a atitude. Para isto é analisada da rede como um todo, utilizando uma visão sociocêntrica (com os nós contendo relações específicas em uma população definida) ou como redes pessoais, em uma visão egocêntrica (com os nós que pessoas específicas possuem, bem como seus grupos pessoais).

Regina Maria (2001) em seu artigo[43] ainda destaca que o acompanhamento das redes sociais oferece a oportunidade de ser feita uma pesquisa qualitativa e quantitativa a respeito dos seguintes pontos:

- De como ela é vista pelos clientes;
- De como está o desempenho dos produtos;
- Das características dos produtos concorrentes;
- Das sugestões de melhoria;
- Sobre os usuários chave e formadores de opinião, aqueles que possuem muitos seguidores;

O ponto principal está na possibilidade de transformar uma enorme quantidade de dados sem sentido em informações produtivas e utilizáveis para o processo de estratégia competitiva, assim facilitando a produção de conhecimentos. Seus produtos gerados podem ser disponíveis em vários formatos, através de gráficos, relatórios e *dashboards*[44] variados.

A ferramenta SCUP possui vários tipos de gráficos onde facilita a leitura e interpretação, no caso de gestores ou de quem toma as decisões sobre ações de marketing ou negócios, pois ilustra todo o cenário das redes sociais que estão sendo

---

[43] http://www.scielo.br/pdf/ci/v30n1/a09v30n1.pdf

[44] O *Dashboard* representa a consolidação de dados, muitas vezes a partir de variadas fontes, e a apresentação dos mesmos de uma forma inteligível e de onde se possam tirar conclusões rapidamente. É basicamente uma forma para os gestores terem um conhecimento rápido dos dados importantes para eles. Além de ser um resumo dos dados importantes, o *Dashboard* também serve como forma de realçar dados específicos, permitindo ao utilizador analisá-los pormenorizadamente.

monitorada. Para ter ideia de exemplo de gráficos a figura 17 representa esses dados:

**Figura 17 – Exemplo de tela de gráfico de resultados do SCUP**

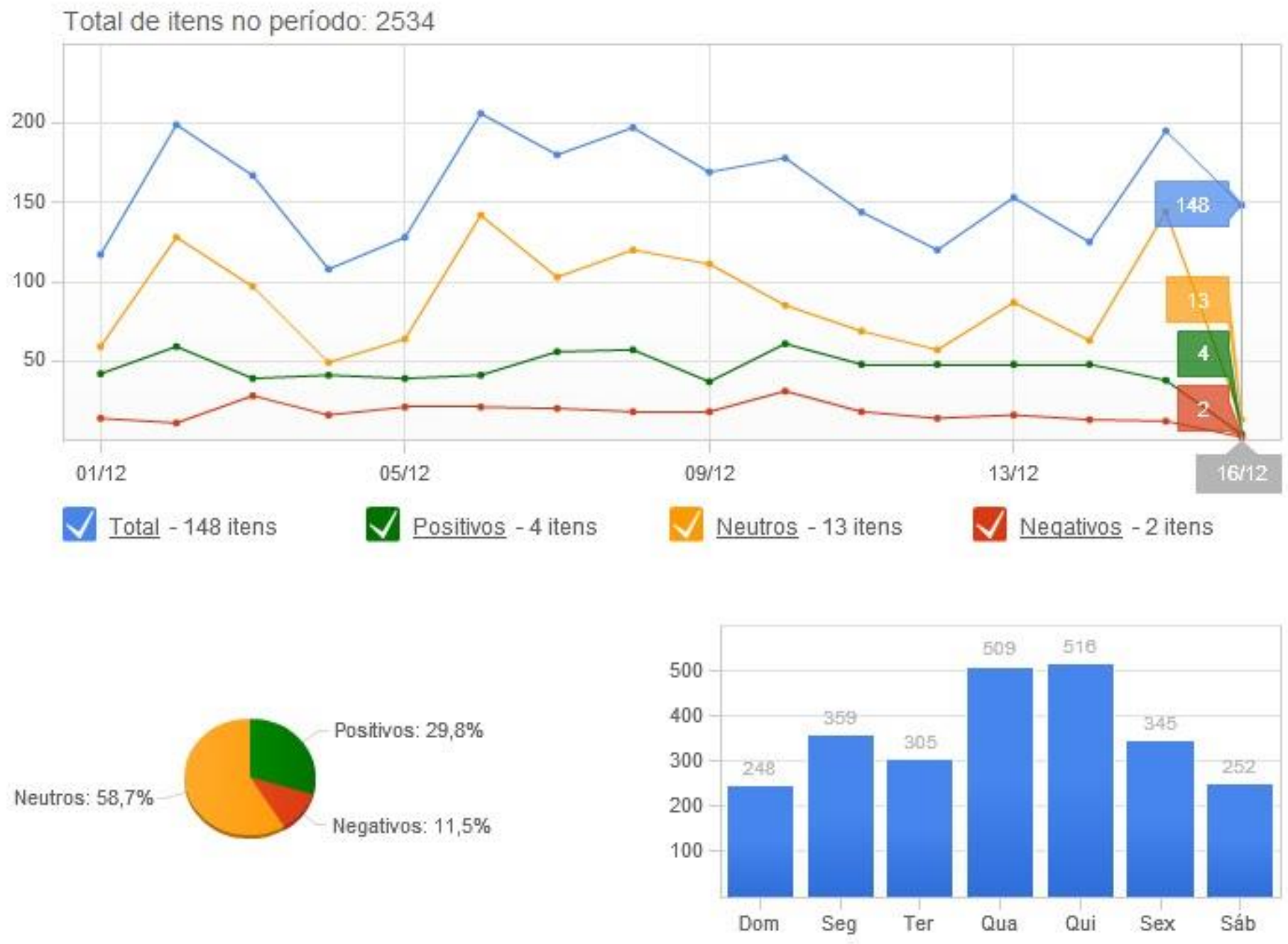

(Fonte: http://4.bp.blogspot.com/_xthNlqAEjAI/TQzmJXCc09I/AAAAAAAAAKQ/KW_C0yWgPF0/s1600/trending.bmp)

Tiago e outros, explica que a “World Wide Web” é composta por uma imensidão de dados que possuem uma estrutura definida ou que não possuem uma semântica explícita. Como exemplo deste cenário, uma página em código HTML possui uma estrutura própria, porém esta não possui informações suficientes. Sendo assim, ao encontrar uma *tag* (linha de código) com a seguinte mensagem “Erro! O nome de arquivo não foi especificado.” em um determinado ponto do arquivo HTML é possível afirmar de que é uma imagem, mas o conhecimento sobre o conteúdo desta imagem é restrito.

A avalanche de dados disponibilizados na Web é tanta que se faz necessário um padrão de estrutura suficiente para a mineração dos dados, como descreve o autor:

> Devido à necessidade de aperfeiçoar os mecanismos de busca e utilizar a Web como plataforma de integração, por meio de serviços, há uma busca crescente em estruturar os dados. Essa estruturação,

> entretanto, deve ter uma flexibilidade dado a própria natureza da Web em que há uma vasta variedade de dados. (FRANÇA, T.C. et. al, 2014)

Mesmo que a *tag* de erro possua um atributo do tipo texto, relativo a essa imagem, textos são dados não estruturados e é necessário aplicar técnicas de mineração de dados para extrair informações desse texto.

Não obstante, isso pode nos enganar. O processo não é simples o bastante ter que analisar uma enorme quantidade de informações que estão soltas e que incluem gírias e ironias em seu interior. Não se deve esquecer que, para a produção de conhecimento, há a necessidade de classificação dos dados obtidos, quanto à veracidade e legitimidade do seu conteúdo e a credibilidade da autoria dessas informações.

# Capítulo IV

## Estudo de caso: a ferramenta SCUP

# 4 A FERRAMENTA SCUP

## 4.1 A história da ferramenta SCUP

SCUP é uma ferramenta e também uma marca, adquirida recentemente, através de transação[45] estratégica de mercado global, pela empresa Sprinklr, norte-americana sediada na cidade de Nova Iorque, empresa que tem como propósito o serviço de monitoramento de mídias sociais de grandes marcas, por meio da busca, classificação, mineração e análise estatística de dados.

A plataforma SCUP foi considerada pela ABCiber - Associação Brasileira de Pesquisadores em Cibercultura como a ferramenta líder no mercado brasileiro de monitoramento on-line em 2012, devido a ferramenta atender ao conjunto de princípios considerados por TUFTE (2006), fundamentais para um bom design referente a análise de dados e informação de modo a maximizar a sua clareza e compreensão evidente do que é representado graficamente:

- Ser fiéis aos dados;
- Ter coerência interna;
- Apresentar os dados em níveis distintos de detalhamento;
- Chamar atenção para o seu conteúdo, e não para a sua construção;
- Incentivar um olhar comparativo acerca dos dados exibidos.

A configuração inicial do sistema pode ser feita em português do Brasil, inglês (americano) ou espanhol. Além disso, a ferramenta também realiza as buscas em mais de uma rede social, permite o resgate de um perfil específico, minimiza o problema do “ponto cego”, isto é, o conjunto de citações relevantes que não são capturados através da busca, pois são protegidas pelo perfil do usuário.

---

[45] http://idgnow.com.br/internet/2015/05/03/sprinklr-compra-scup-com-foco-em-expandir-monitoria-de-redes-sociais/

O SCUP monitora os canais: *Twitter*, *Google* notícias, alertas e *blogs*, *Facebook*, *YouTube*, *Flickr*, *Yahoo* Respostas, Reclame Aqui, *WordPress Feed*, Orkut (comunidades cadastradas), RSS (feed), *SlideShare* e *Vimeo*.

Mais detalhes sobre a ferramenta podem ser visualizados nos anexos 1, 2, 3, 4 e 5.

## 4.2 Detalhes da ferramenta

O SCUP baseia-se em quatro módulos: monitor, relacionamento, publicação e relatórios. Através do módulo Monitor, o usuário pode acompanhar os comentários nos diversos canais em tempo real e classificá-los como positivos, neutros e negativos, além de opção de marcar os comentários por termos mais utilizados e relevantes para sua marca.

Existem várias opções de filtros de visualização como buscas por canal, período, termo, classificação, marcação, etc. Este tipo de recurso auxilia o processo de durante a classificação.

No módulo relacionamento, a partir das buscas cadastradas, a ferramenta apresenta a lista dos usuários que interagiram nas mídias sociais. Essa lista permite que o usuário agrupe perfis e faça anotações sobre evangelizadores e destruidores da sua marca. Tudo é uma questão de configurar a ferramenta de acordo com a necessidade das informações.

O módulo de publicação do SCUP é bastante simples, podendo agendar seus *posts* no *Twitter*, *Facebook* e *Youtube* e ainda pode-se utilizar o recurso de encurtar URL via Scup.it, Bit.ly ou Gloo.gl. A ferramenta também oferece a opção de cadastrar diferentes perfis de colaboradores: visitante, gerenciador, administrador, classificador, moderador e publicador.

No último módulo, o de relatórios, o SCUP gera extratos por classificação, marcação, abrangência, interações e comparativo de monitoramentos. O relatório por termos deixa um pouco a desejar, pois gera um consolidado e deixa a cargo do analista a confecção de relatórios de termos por volume de menções, por canal e classificação.

Todos os dados podem ser exportados em arquivo .csv, desta forma é possível personalizar os relatórios da maneira que quiser.

Como na maioria das ferramentas de monitoramento pleno, o SCUP utiliza as API´s (*Application Programming Interface* ou em português, Interface de Programação de Aplicativos) disponibilizadas de cada mídia social para a coleta de registros. Cada serviço ou mídia social monitorada possui suas próprias características em relação ao funcionamento de seu mecanismo de busca, por isso é importante conhece-los para atingir o melhor resultado esperado.

Existem algumas variações e é possível cadastrar uma expressão comum (palavra-chave) para todos os serviços, porém é necessário tomar cuidado com certos operadores (caracteres especiais) para que a busca não traga resultados inesperados.

No SCUP as regras podem ser definidas como ações automáticas que o sistema realiza a partir de critérios pré-cadastrados pelo usuário da ferramenta. Por exemplo: Se você definir previamente que todo item que contiver a palavra “bom” deve ser classificado como positivo, o sistema o fará de modo automático.

É permitido fazer um comparativo de frequência em cada busca cadastrada que pode trazer diversos resultados e com isso é possível, entender qual mídia (*Facebook*, *Twitter*, Reclame aqui entre outras) tem mais representatividade para a empresa ou marca que está sendo monitorada, podendo compará-las em abrangência. Pode ser acompanhado o número de publicações, tempo médio de resposta, respostas por faixa de tempo e filtrar esses dados para identificar a eficiência e frequência de postagens nas redes. Dessa maneira, são disponibilizados dados preciosos em relação a uma operação de atendimento em mídias sociais.

### 4.2.1 Vantagens do SCUP:

Abaixo são pontuadas algumas das principais vantagens da ferramenta de monitoramento e seus detalhes (de acordo com o fabricante):

- Duas ou mais pessoas podem utilizar uma conta ao mesmo tempo para monitorar permitindo o trabalho de uma equipe de analistas, podendo compartilhar cada um dos seus monitoramentos com diversos outros colaboradores. O acesso é independente por monitoramento, o que traz ainda mais flexibilidade para montar sua estrutura de acesso. Cada colaborador pode estar em um dos vários perfis de acesso (Administrador, Gerenciador, Moderador, Publicador, Classificador, Visitante), que permitem organizar o processo de seu dia-a-dia no SCUP, ajudando a garantir o nível de segurança que necessite. A postagem para o *Twitter* pode passar por aprovação. Junto com o sistema de perfis de acesso, é possível definir usuários que podem apenas visualizar, postar e aprovar publicações na rede. Um processo simples e eficiente.
- Apresenta a frequência e o valor do sentimento (positivo, neutro e negativo) das palavras mais citadas, permitindo que a valoração seja colocada pelo analista.
- Oferece um plano acadêmico que possibilita alunos e professores tenham acesso aos recursos da ferramenta por um período de dois meses.
- Possui um blog onde podem ser discutidas novidades e dúvidas sobre a ferramenta.
- Permite exportações completas para Excel dos dados de Mídias Sociais e com isso esses dados podem ser consumidos e integrados a outros sistemas, além de poderem ser analisados em plataforma externas para análises específicas dentro de certos contextos, além de possuir uma API aberta, para exportar dados dos monitoramentos e integrá-los com plataformas externas.
- Possibilita o monitoramento no site Reclame Aqui, um dos sites nacionais mais conhecidos na divulgação da insatisfação do cliente.

### 4.2.2 Desvantagens do SCUP:

Da mesma forma, também existem alguns pontos que podem não auxiliar em processos específicos na hora da implantação:

- Não automatiza o processo de categorização necessitando que seja realizado por um Analista;
- O cruzamento entre monitoramentos e *tags* é limitado;
- Não possui uma boa coleta de dados do *Facebook* e Orkut, tendo mais destaque as coletas provenientes do *Twitter*.

## 4.3 API do SCUP

Além de todas as funcionalidades do SCUP, outra função importante da ferramenta é sua API que possibilita aos desenvolvedores integrar as funcionalidades do SCUP com seus sistemas.

Para que os desenvolvedores possam obter os dados do SCUP basicamente precisam desenvolver uma aplicação que conecte com a API e interprete o formato de resposta JSON. A API do SCUP é aberta, utiliza-se do princípio REST com formato de resposta JSON. Um exemplo de funcionamento da API pode ser vista na figura 18 abaixo:

**Figura 18 – A API do SCUP**

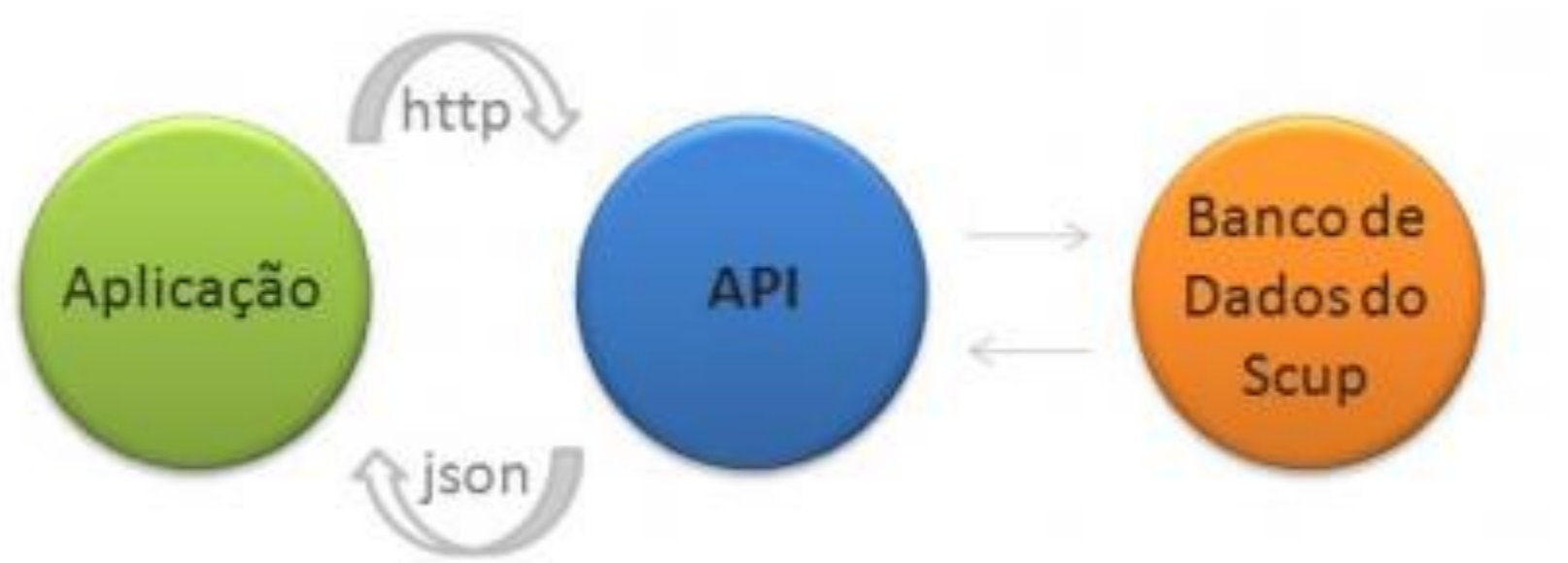

(Fonte: http://blog.scup.com.br/o-scup-agora-tem-api)

### 4.3.1 Métodos

Esses são os métodos disponíveis na API do SCUP e todos precisam de autenticação para obter reposta. São eles:

- Listagem de Monitoramentos;
- Listagem de Buscas;
- Listagem de Itens;
- Listagem de Usuários;
- Listagem de Itens do Usuário;
- Tagueamento;
- Classificação;
- Responder um Item;
- Apagar Itens;
- Logs de Itens;
- Resposta de um item.

A API possui limite de chamadas por hora, caso este limite seja atingido, a seguinte mensagem será retornada: “atingido limite de chamadas”. Quando o ID do monitoramento informado não existe ou você não tem permissão para acessar os dados do monitoramento, a seguinte mensagem será retornada: “monitoramento não encontrado”.

A API possui um limite de 15 páginas para as requisições realizadas. Se precisar utilizar mais de 15 páginas para trazer tudo que deseja, então será necessário restringir a sua busca, por exemplo, modificando a data procurada.

## 4.4 Casos de monitoramento

O uso da ferramenta SCUP proporciona um auxílio em processos de acompanhamento do que seus clientes dizem nas redes sociais e também como forma de interação e ouvidoria dos consumidores em casos de dúvida ou problemas.

Os dois casos a seguir relatam o quanto a ferramenta pôde ajudar nesses pontos e quais os retornos que as empresas obtiveram ao escolher essa ferramenta diante de um mercado com algumas opções disponíveis. O texto foi retirado da publicação do site SCUP – Cases[46], pois são relatos dos clientes que utilizaram a ferramenta e contribuíram para que os autores produzissem o texto. Vejamos com atenção.

### 4.4.1 Caso TAM[47]

**Os primeiros passos do SAC 2.0 da TAM**

A trajetória da TAM sempre foi guiada pela disponibilidade de servir seus clientes; nas redes sociais, essa filosofia não poderia ser diferente. Desde 2010 atuando nesses canais, a empresa busca marcar sua presença digital pela prestação de serviços e atendimento ao usuário. "Queremos ajudar quem escolhe voar com a gente em todos os momentos", diz Ester Bonança, coordenadora FALE 2.0 da TAM Linhas Aéreas.

A entrada da empresa no *Twitter* ocorreu por uma iniciativa do time de marketing, auxiliado por uma agência externa. Em pouco tempo, as interações dos usuários cresceram fortemente e a área precisou envolver outros departamentos da empresa para atender às demandas do SAC 2.0.

No final de 2010 a operação começou a se profissionalizar. Com o contínuo crescimento da operação, o atendimento ao cliente nas redes sociais passou a ser administrado diretamente pelo "Fale com a Gente", o que marcou o início de uma operação estruturada de SAC 2.0 na TAM, com uma equipe interna, bem preparada e capacitada para lidar com as mais diversas situações.

**Atendimento global com rapidez e eficiência**

46 https://www.scup.com/pt/cases/
47 SCUP Case produzido por SCUP S/A em parceria com TAM Linhas Aéreas em 21/02/2014;

Com operação em diversos fusos, voos praticamente 24 horas por dia e clientes espalhados pelo globo, os desafios da TAM nas redes sociais vão muito além do imediatismo do dia a dia.

A empresa trabalha com a pressa da indústria aérea, compromissos e agendas apertadas, sempre com foco na segurança. “Nas redes sociais, os prazos são realmente muito curtos, tanto para resolver uma insatisfação como para dar uma informação”, diz Ester. Dessa forma, a TAM precisa estar sempre pronta para responder o mais rápido possível. Para isso, precisa coletar todas as menções e postagens rapidamente e responder as dúvidas de forma controlada e organizada.

A empresa usava uma ferramenta para redes sociais que apresentava limitações. “A nossa principal dificuldade era a falta de suporte no Brasil, tanto para *help-desk* quanto para nos ensinar como operar com a ferramenta, que era em outro idioma e menos amigável”, lembra a coordenadora. “Era essencialmente uma ferramenta de métricas adaptada para ser usada no atendimento”.

**Os benefícios do SCUP para a operação da TAM**

No meio de 2012, depois de três meses de pesquisa de mercado, a TAM conheceu o SCUP. “Eu acredito que na época era a única ferramenta voltada para atendimento em si. E isso chamou a nossa atenção”, lembra Ester. “Com o SCUP, podemos fazer tudo num só painel, com muita praticidade e sem precisar customizar. Além disso, a ferramenta está em português, possui suporte no Brasil, com um SLA de atendimento muito bom. Sem contar que a equipe nos ajuda a otimizar nosso monitoramento, permitindo que ela seja usada de forma rápida e eficaz”.

Continua Ester: “O SCUP permite que a gente consiga capturar todas as menções, seja o nosso perfil citado ou não, se organizar para trabalhar, responder de forma rápida e precisa, estar presente para o nosso cliente no momento em que ele postar. O que diferencia o SCUP é a organização. No painel, você responde, ‘retuíta’, segue, classifica por tipo e sentimento e consegue filtrar da forma que você quer”.

Os cinco analistas de atendimento sob a coordenação de Ester vieram de diversas áreas da TAM, já trazendo consigo um forte conhecimento sobre a operação, a cultura da empresa e o segmento de aviação. Durante sete dias por semana, eles prestam atendimento em três idiomas, ajudando os clientes com reclamações, sugestões, elogios e assuntos mais corriqueiros, como dúvidas de atualização de reserva, confirmação de assento, pedidos de horário de voo, além das curiosidades sobre a empresa e o mundo da aviação.

Desde o momento em que o SCUP coleta um post, a equipe faz uma classificação por assunto e sentimento, além de buscar o histórico de antigas interações do cliente. “Com o armazenamento de histórico do SCUP, temos que pedir os dados para o cliente só uma vez. Isso já fica guardado para uma próxima interação. É muito difícil encontrar uma ferramenta que faça esse armazenamento de forma concisa como o SCUP consegue fazer”, afirma Ester.

Para ela, outro diferencial do SCUP é a abertura de tickets de atendimento, que permite saber se o cliente já está sendo assistido.

“Como as redes sociais não permitem uma burocratização ou formalidade nós temos mais condições de prestar informações ao nosso cliente sobre o status de seu processo. Percebemos que assim as pessoas começam a conhecer melhor a nossa empresa, se sentem acolhidas e conseguem ver o nosso trabalho. Esse é o verdadeiro relacionamento”, afirma Ester. “A principal diferença do SAC 2.0 é a facilidade para criar o relacionamento” Ester Bonança, Coordenadora FALE 2.0 da TAM Linhas Aéreas.

**A construção de relacionamentos duradouros e genuínos**

Quatro anos após o ingresso nas redes sociais, mais de um ano usando o SCUP, uma equipe dedicada e apaixonada pelo que faz, a TAM registra um aumento de 90% nos atendimentos na comparação com a antiga ferramenta, atingindo uma média de 300 interações por dia. “Hoje somos uma das empresas que mais interagem no *Facebook* e o SCUP nos ajuda muito nessa tarefa”, diz Ester Bonança.

“É importante promover a marca, explicar a filosofia e o serviço que a gente busca oferecer, mas a melhor forma de divulgar um bom serviço são os nossos clientes aprovando o que fazemos, interagindo e criando um vínculo forte. O que resume a TAM no 2.0 é o espírito de servir, que está em nosso DNA”.

Seguindo essa filosofia, a empresa busca fornecer a primeira resposta em até duas horas depois da publicação, respondendo a maioria das interações entre 15 e 20 minutos.

Para a TAM, “só as empresas que criarem um relacionamento de confiança com seus clientes garantirão a lealdade e o crescimento de seus negócios.”

**Como o SCUP ajudou a TAM?**

- ✓ Crescimento do número de atendimentos em 90%;
- ✓ Atendimento das 8h às 20h, 7 dias por semana, com múltiplos operadores;
- ✓ Armazenamento do histórico completo de interações.

**Por que a TAM confia no SCUP?**

- ✓ Suporte em língua portuguesa;
- ✓ Treinamento e capacitação para uso da ferramenta;
- ✓ Interface e funcionalidades específicas para atendimento;
- ✓ Captura de menções abrangentes e em tempo real;
- ✓ Classificação por tipo e sentimento;
- ✓ Filtros de menções;
- ✓ Possibilidade de responder e seguir usuários;
- ✓ Centralização da operação de atendimento num só lugar;
- ✓ Acesso a todas as interações e dados de usuários;
- ✓ Abertura de *tickets* de atendimento.

### 4.4.2 Caso NETSHOES[48]

**História**

Com 14 anos de existência, a Netshoes, que foi primeira loja virtual de artigos esportivos da América Latina e é hoje o maior "*e-commerce pure play*" de artigos esportivos do mundo, alcançou o surpreendente faturamento de mais de R$ 100 milhões com as redes sociais. O resultado vem de um forte investimento em marketing e conteúdo estratégico para as redes sociais, aliado a uma visão que coloca a Central de Relacionamento como carro-chefe da organização.

A empresa, que foi uma das primeiras a atender seus clientes através dos canais sociais no Brasil, conta com o SCUP para trabalhar nas redes sociais desde 2011. E em 2012, com a expansão para a Argentina e México, o uso da ferramenta se estendeu para América Latina.

**Desafios**

Antes de maio de 2011, o atendimento nas mídias sociais na empresa, referência em e-commerce no Brasil, encontrava três limitações básicas. A primeira era a descontinuidade do suporte: um cliente que buscava ajuda pela segunda ou terceira vez, por exemplo, precisava contar novamente toda a sua história a cada atendimento. Além disso, o monitoramento de redes sociais era apoiado por plataformas que traziam muito "lixo", o que dificultava saber qual era realmente a imagem que a marca projetava na cabeça dos seus consumidores. Finalmente, sem conhecer os números de suas próprias operações, a empresa enfrentava ainda um terceiro obstáculo: a impossibilidade de medir e visualizar resultados.

"O SCUP nos permite organizar um histórico dos diálogos que mantemos nas redes. Isso nos faz economizar tempo e melhorar o atendimento que oferecemos" Beatriz Rosado, da Netshoes.

---

[48] Case produzido pela Equipe SCUP em parceria com Netshoes em 29/07/2014;

**Solução**

A contratação do SCUP desatou diversas amarras às quais a área estava submetida. Em primeiro lugar, a manutenção de uma sequência contínua da conversa com o cliente deixou de ser um desafio. “O SCUP nos permite organizar um histórico dos diálogos que mantemos nas redes”, afirma Beatriz Rosado, uma das analistas responsáveis pelo atendimento da Netshoes nas mídias sociais. “Isso nos faz economizar tempo e melhorar o atendimento que oferecemos”.

Outra dificuldade enfrentada pela Netshoes no passado era a falta de qualidade dos dados oferecidos por outros serviços. A equipe, que hoje usa o SCUP todos os dias, já na fase de teste sentiu a diferença na pertinência dos dados trazidos pela plataforma. “Com o SCUP passamos a capturar menções que falam direta e efetivamente de Netshoes, algo que fez toda a diferença”, diz a analista Andressa Jordano.

A área passou a mensurar seus resultados de forma consistente. Beatriz lembra que, antes do SCUP, não era feito nenhum relatório sobre o trabalho de relacionamento. “Agora, usamos a ferramenta não apenas para fazer atendimento e monitoramento, mas também para gerar relatórios de performance”, explica. “Tudo que tem a ver com números, quantas interações, quantos compartilhamentos, nós extraímos do SCUP”.

O trabalho se tornou não apenas mais simples, mas também mais relevante para o negócio. “Antes existia uma prática de ‘atender e pronto’”, afirma Andressa. “Mas e os clientes que não estamos atendendo diretamente, por exemplo? O que é a Netshoes para eles? Essa é uma demanda muito importante que o SCUP atendeu”.

“Extraímos do SCUP insights que muitas vezes podem corroborar ou validar algum sentimento que temos em relação a alguma ação.” Douglas Costa, Netshoes

Mais tarde, o serviço começou a ser usado não apenas pela área de relacionamento, mas também pelo departamento de marketing e comunicação, que constrói análises e traça estratégias com os dados gerados pelo monitoramento.

“Extraímos do SCUP insights que muitas vezes podem corroborar ou validar algum sentimento que temos em relação a alguma ação”, afirma Douglas Costa, gerente de Marketing & Comunicação da Netshoes.

Em 2013, com o objetivo de se superar para maximizar as vendas e adequar a estrutura à demanda, por meio de iniciativas criativas e da atenção aos pequenos detalhes, a Netshoes usou como estratégias algo que até então passou despercebido, mas que hoje se tornou fundamental para seu sucesso - posicionar a Central de relacionamento como carro-chefe da organização, conta Juliana Pires, Gerente Geral da Central de Relacionamento da Netshoes.

Com uma metodologia simples, a equipe de Planejamento Estratégico ganhou a responsabilidade de fechar as parcerias com todas as outras áreas da empresa, como Tecnologia da Informação, Marketing e Financeira. O critério de avaliação de cada etapa está diretamente ligado aos resultados, seja de aumento do percentual de atendimento e eliminação de fila de espera, nos quais contamos com o SCUP para mensurar no SAC 2.0, ou até mesmo, o crescimento do ticket médio de vendas e satisfação dos clientes, complementa Juliana.

“O SCUP faz com que seja um processo fácil e objetivo, trazendo os resultados de que precisamos. Com isso, conseguimos otimizar o nosso trabalho e ganhar tempo.” Beatriz Rosado , Netshoes.

Em julho de 2014, a empresa atribuiu mais de R$ 100 milhões de seu faturamento às redes sociais. Com mais de nove milhões de seguidores em todas as redes, a Netshoes registra uma média mensal de 52 mil interações com os usuários, entre diálogo, curtidas e compartilhamentos, mostrando a relevância que a Central de Relacionamento e o trabalho nas redes sociais trazem para empresa.

Com o SCUP, a Netshoes conseguiu direcionar o tempo das redes sociais para o que realmente importa - olhar para o cliente. “O SCUP faz com que seja um processo fácil e objetivo, trazendo os resultados de que precisamos”, diz Beatriz. “Com isso, conseguimos otimizar o nosso trabalho e ganhar tempo. Foi um grande passo para a nossa área e para a Netshoes como um todo”.

Com tamanho sucesso da operação, a empresa também passou a contar com o SCUP no atendimento ao consumidor na Argentina e no México.

**Como o SCUP ajudou a Netshoes?**

- ✓ Organização do histórico de atendimentos nas mídias sociais;
- ✓ A equipe gasta entre 71% e 90% menos tempo com a classificação das mensagens coletadas nas mídias sociais;
- ✓ Geração de relatório de performance.

# Capítulo V

# Conclusão

# 5 CONCLUSÃO

As redes sociais são um caminho sem volta. Elas transpuseram o uso meramente de lazer para adquirir um caráter sociológico, de gerador de identidades coletivas. Dentro dos ambientes virtuais, o conhecimento, a informação, as marcas passam a serem criadas e geridas coletivamente e de forma colaborativa, democrática, interativa, customizável e sem limites de espaço e tempo (BACCI e LOUVERT, 2009).

Nestes ambientes, mesmo com perfis corporativos, o poder é total do usuário. Assim podemos perceber que o monitoramento de redes sociais deixou de ser tendência e virou necessidade. Onde não se busca saber o sentimento de um cliente pela marca, mas sim sua percepção de compra e uso, que vão ser transformados em dados com diversas aplicações interessantes, devido ao grande número de dados disponibilizados pelos usuários involuntariamente.

Redes como Blogs, Fóruns, *Twitter*, *Facebook*, *Flickr*, *YouTube* e vários outros ambientes sociais são um poço de informações sobre o que os usuários da Web estão interessados e quais são as suas opiniões sobre um assunto qualquer.

Apesar dos desafios, a realidade cada vez mais, mostra que as empresas que já aceitaram o monitoramento como uma realidade estão conseguindo minimizar danos ao mesmo tempo em que auxilia empresas nas diversas tomadas de decisão do cotidiano.

Muitos desafios a serem enfrentados ainda são existentes, entretanto a possível forma de se trabalhar com amostras mais volumosas de dados das redes sociais online permite que novas informações sejam sacadas e outras sejam mais consistentes, tendo em vista que a amostra será em um universo maior.

Além do desafio técnico de averiguar enormes quantidades de dados, outros desafios nascem a partir desta inovadora oportunidade de análise, visto que, novas informações que em um momento anterior eram descartadas, agora são extremamente consideráveis.

Dar tratamento a essas informações novas de forma adequada ultrapassa as áreas técnicas da computação, e porque não das ciências exatas, visto que outros conhecimentos como o de humanas (antropologia, sociologia, psicologia, entre outros) são úteis nesse momento.

Esta pesquisa introduziu o assunto de "*Big data"* e a análise das redes sociais dando permissão que outros pesquisadores que pretendam trabalhar com grandes bancos de dados conheçam os principais tópicos e desafios que existem atualmente.

Ao mesmo tempo, os profissionais que trabalham com o conceito de "*Big data*" podem agora estender seus trabalhos para a análise de dados sociais digitais, que se trata das mídias sociais como um todo.

## 5.1 Resultados

As redes sociais online se tornaram muito populares e parte do nosso dia, exigindo nossa atenção, provocando o surgimento de uma nova avalanche de aplicações disponíveis na Web. A cada dia, a cada segundo, enormes quantidades de dados de conteúdo compartilhados, distribuídos e milhões de usuários interagindo através de nós sociais. Não bastante de tanta popularidade, a análise das redes sociais ainda está no seu início, já que estes ambientes estão ainda experimentando novas tendências e enfrentando variados e oportunos problemas e questões adversas.

Mídias sociais compõem situações perfeitas para o estudo de vários temas da computação, incluindo sistemas multimídia e a interatividade homem-máquina. Além do mais, por aceitar que usuários criem novas informações a todo o momento, aplicativos e ferramentas sobre redes sociais vêm ganhando espaço em pesquisas relacionadas à organização e tratamento de dados em grandes quantidades, além de construírem um ambiente perfeito para a extração de conhecimento e domínio e também aplicações de mineração e tratamento de dados.

Os dados indicam que é um mercado com um futuro a ser explorado e com muito espaço para trabalhar e crescer. As organizações que adotarem as redes sociais como ferramenta de auxílio às estratégias de negócio, para se comunicar com seus clientes, por exemplo, estão crescendo mais do que as que não fazem o uso dessas ferramentas.

À medida que os meios de comunicação vão evoluindo, os mecanismos de interatividade com o público-alvo devem também passar por um processo de modernização, acompanhando essa nova tendência.

Desta maneira, interagir com o cliente e criar novas oportunidades de entender suas vontades e preferências em tempo real, se torna um ótimo negócio que irá se transformar em performance nos lucros da empresa.

## 5.2 Trabalhos futuros

O que esta pesquisa quer deixar para uma futura sequência de trabalhos está em torno das redes sociais digitais, e sobre o ambiente de dados Web que possibilita ser explorado e analisado como vantagem competitiva empresarial.

Desta forma, acompanhando os raciocínios dos autores citados neste trabalho, devemos nos preocupar em questões sobre os dados sociais e como eles provocam o mercado e regulam o ambiente de especulação empresarial, dando ênfase aos assuntos mais tratados na rede. Isso sustenta a teoria de monitoramento e análise das mídias digitais e o que esse recurso auxilia na gestão empresarial.

A internet das coisas é algo presente em grandes centros urbanos e desta forma, a pesquisa deixa em aberto discussões sobre esse universo digital, onde as mídias sociais e as “coisas” podem interagir e gerar informações preciosas para empresas que olhem para os dados de forma diferente que seus concorrentes.

As redes sociais específicas para cada tipo de usuário ou grupo social digital faz com que o monitoramento seja imprescindível para empresas que desejam focar seus produtos e serviços daqui pra frente.

A internet possibilita um trabalho imaginário de ideias novas a todo o momento. O usuário é o maior fornecedor de dados e sempre querendo o retorno dessas informações em forma de ofertas, qualidade, e interatividade que lhe agrade na rede. Assim, quem estiver mais bem preparado para essa visão diferenciada estará alinhado aos gostos e preferências da sociedade digital.

## REFERÊNCIAS

# ANEXOS

## Anexo 1 – Tela inicial do Sistema SCUP

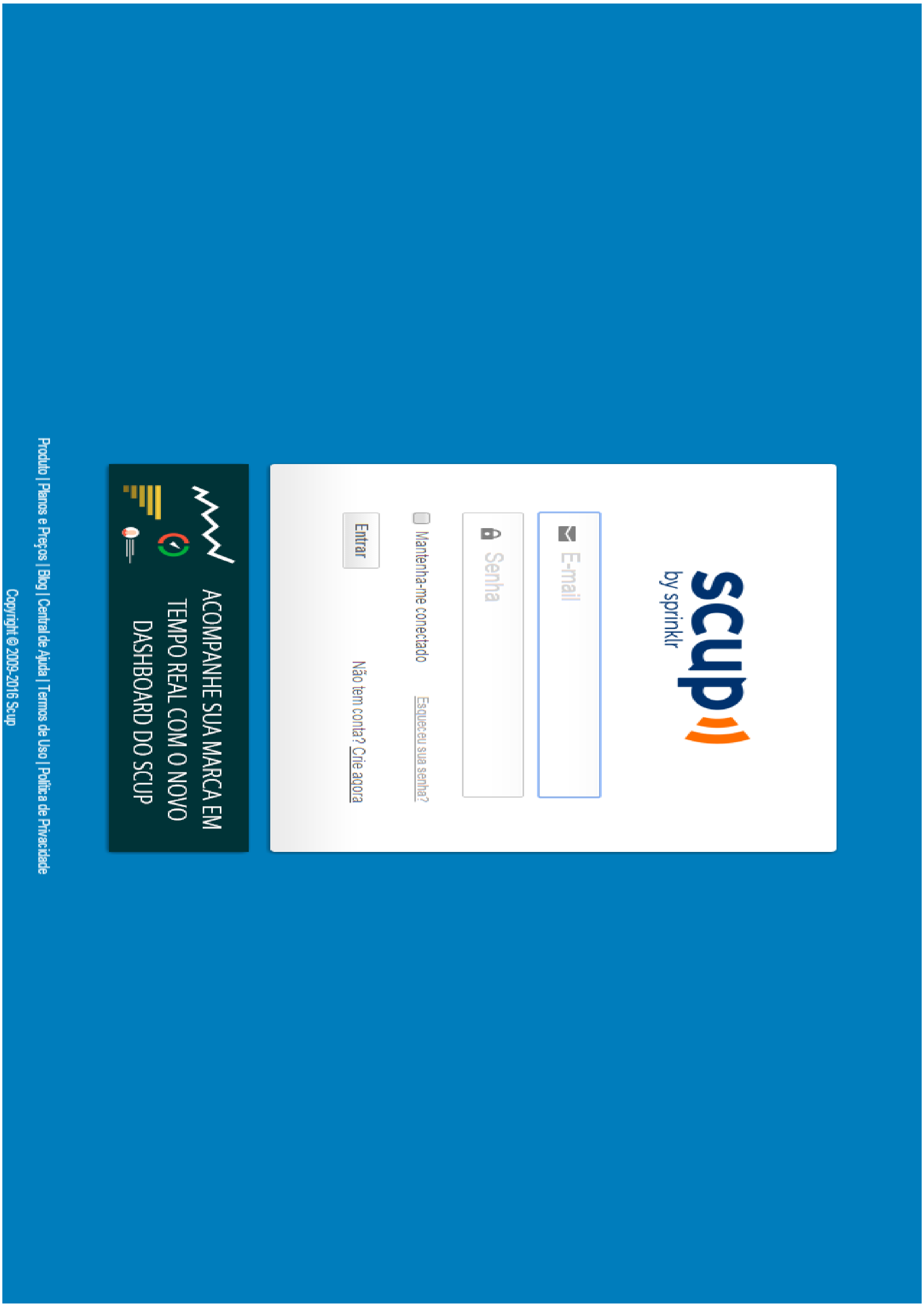

## Anexo 2 – Tela inicial da conta do usuário

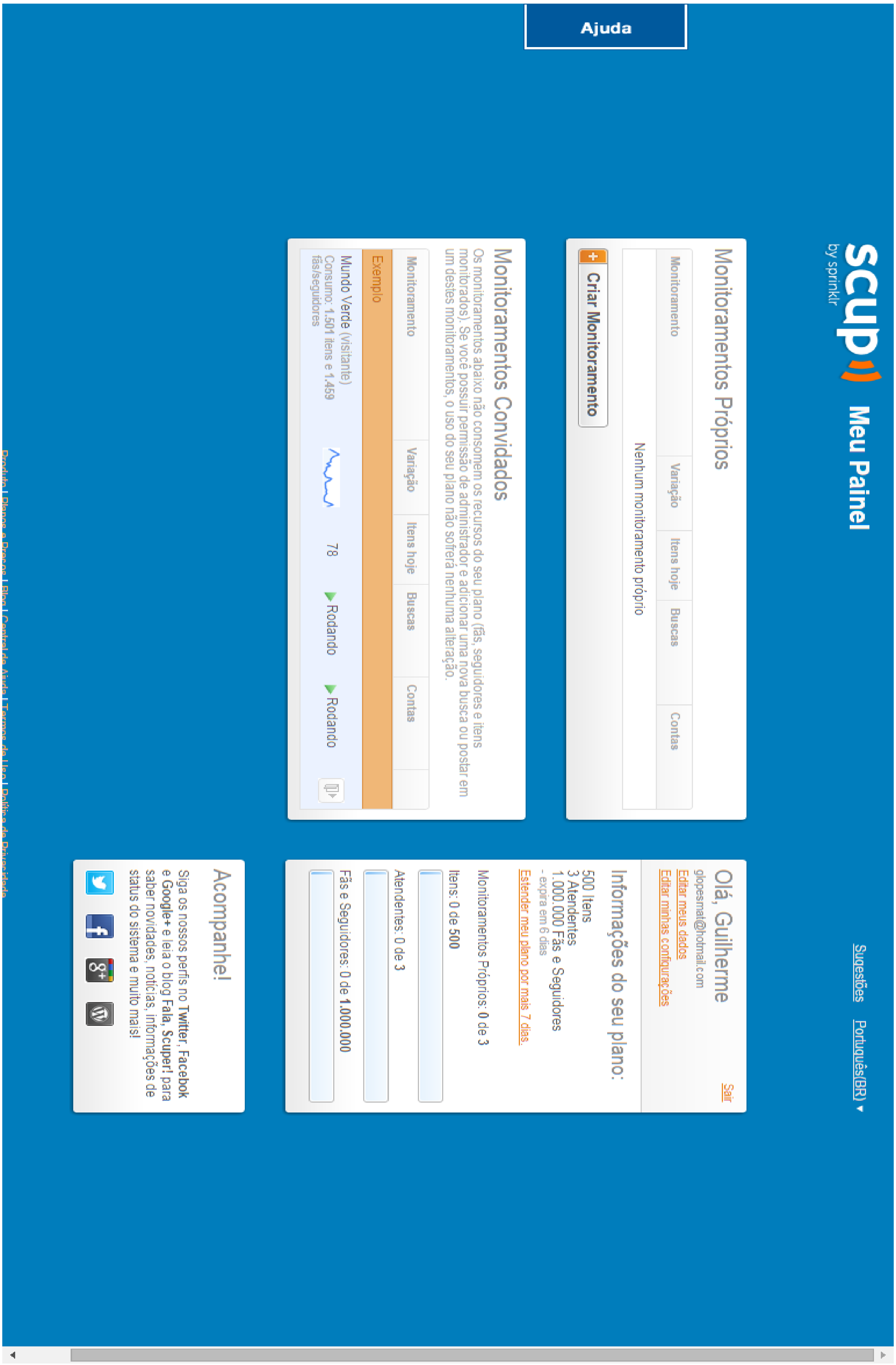

# Anexo 3 – Processo de criação de monitoramento

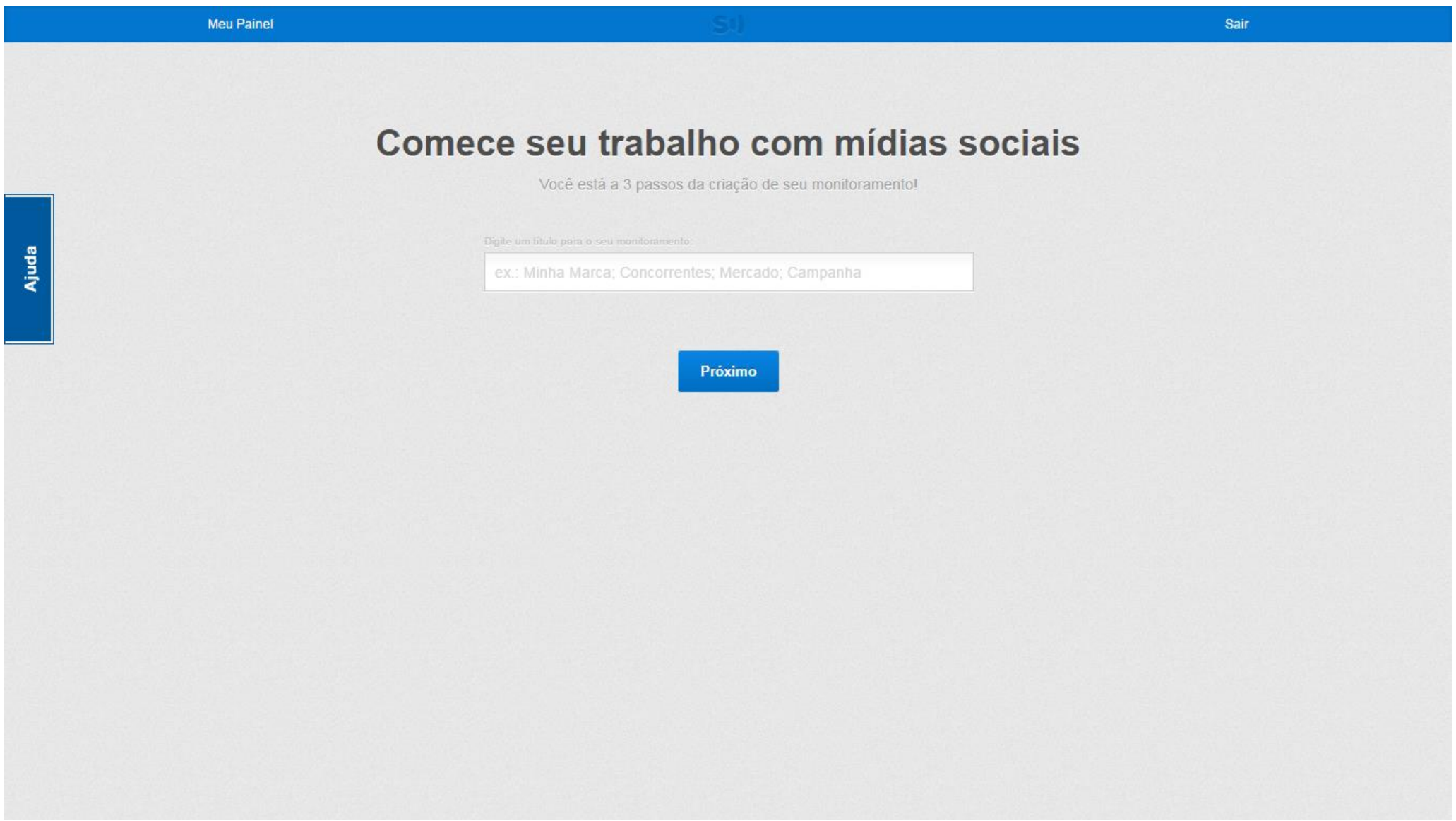

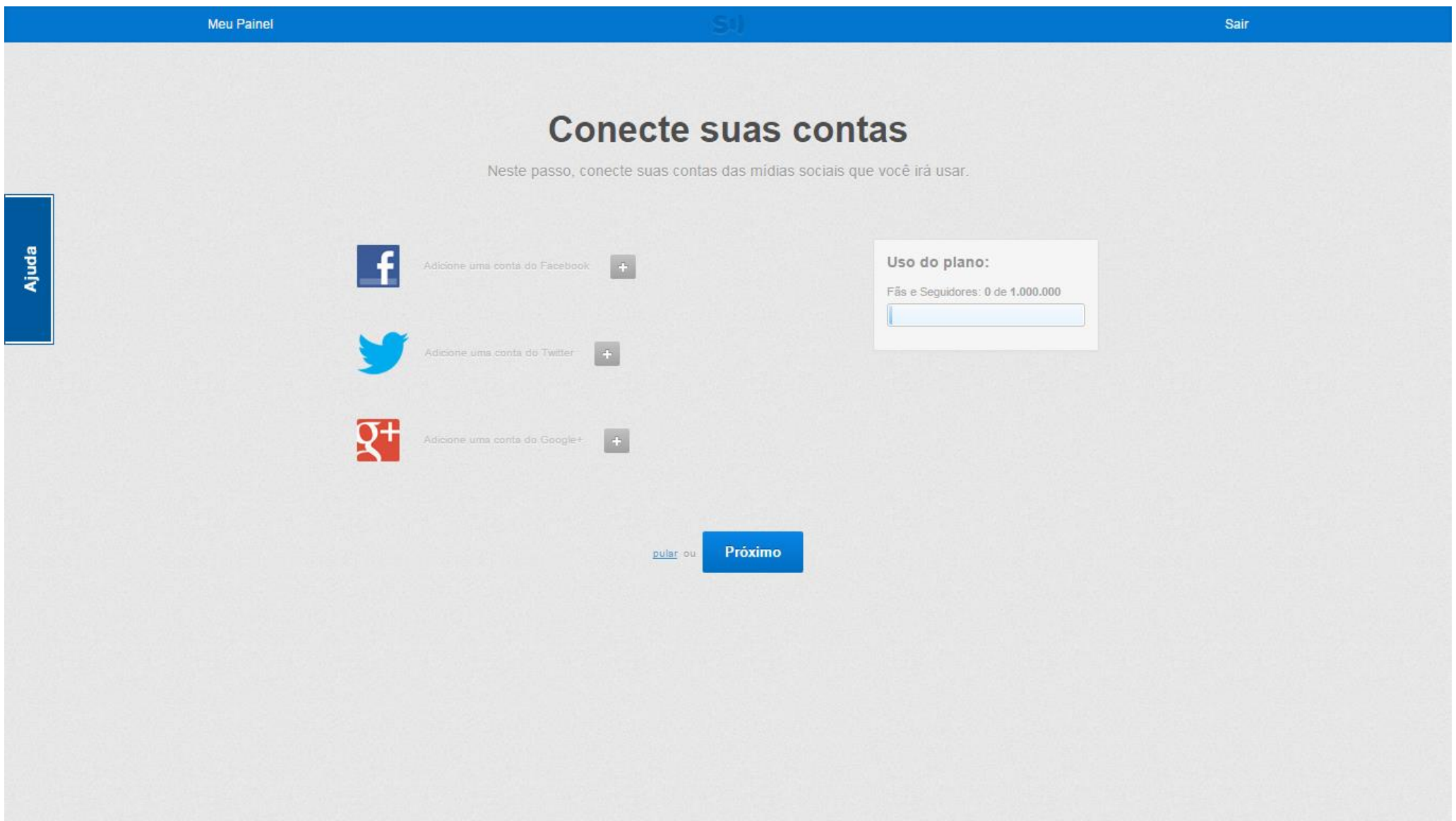

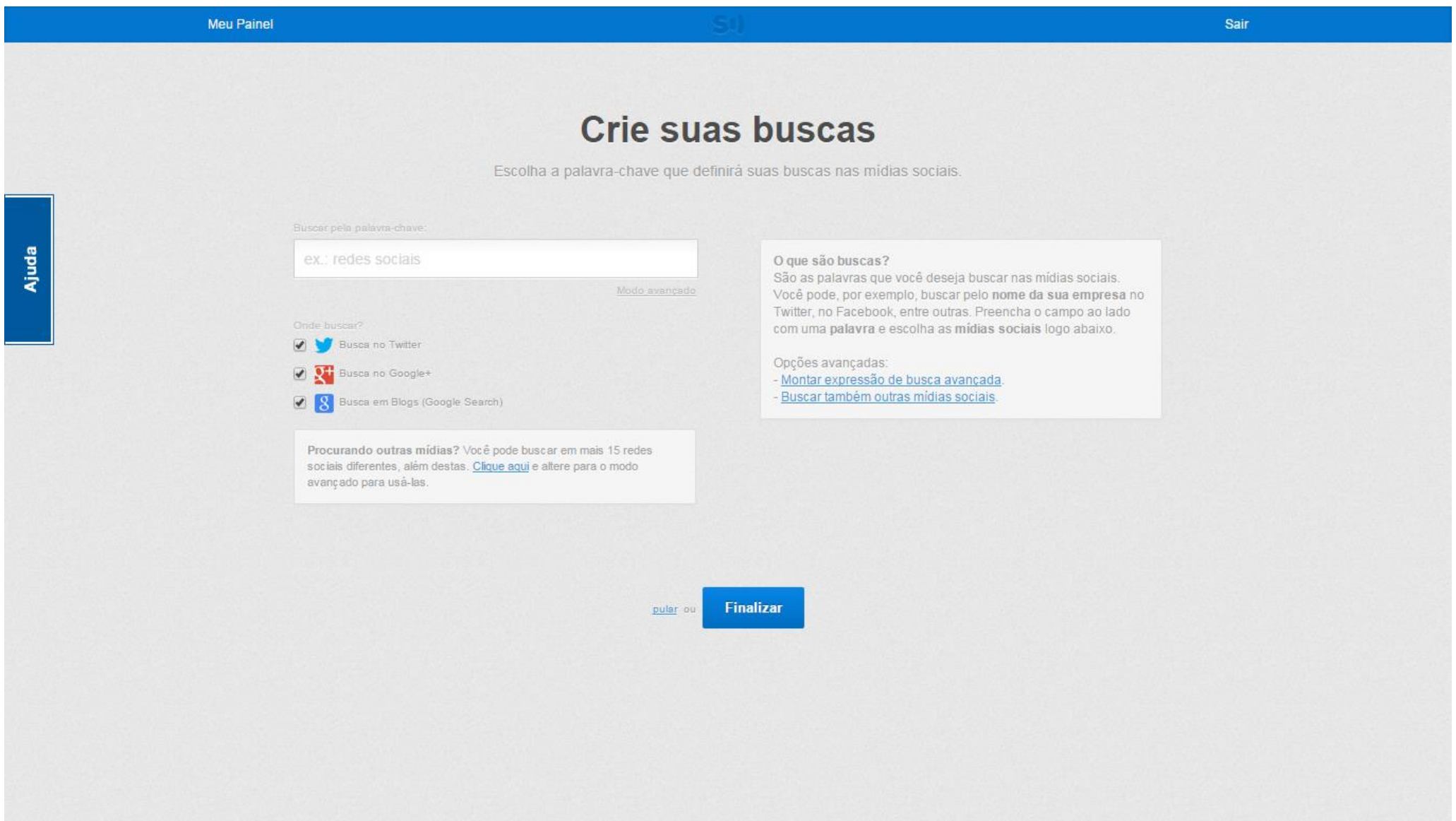

## Anexo 4 – Tela dos planos e preços do SCUP

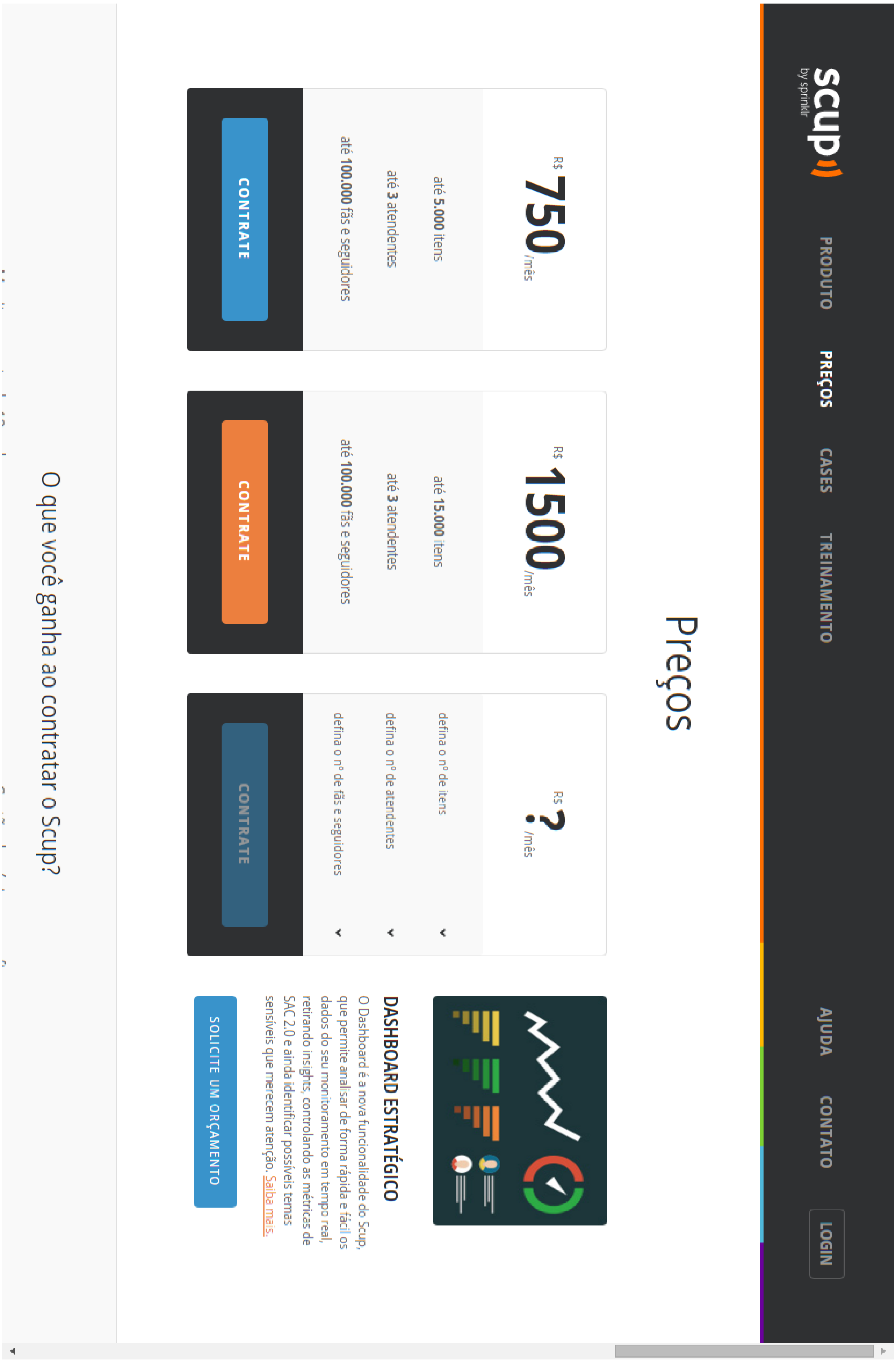

**Anexo 5 – Tela de Filtros do SCUP**

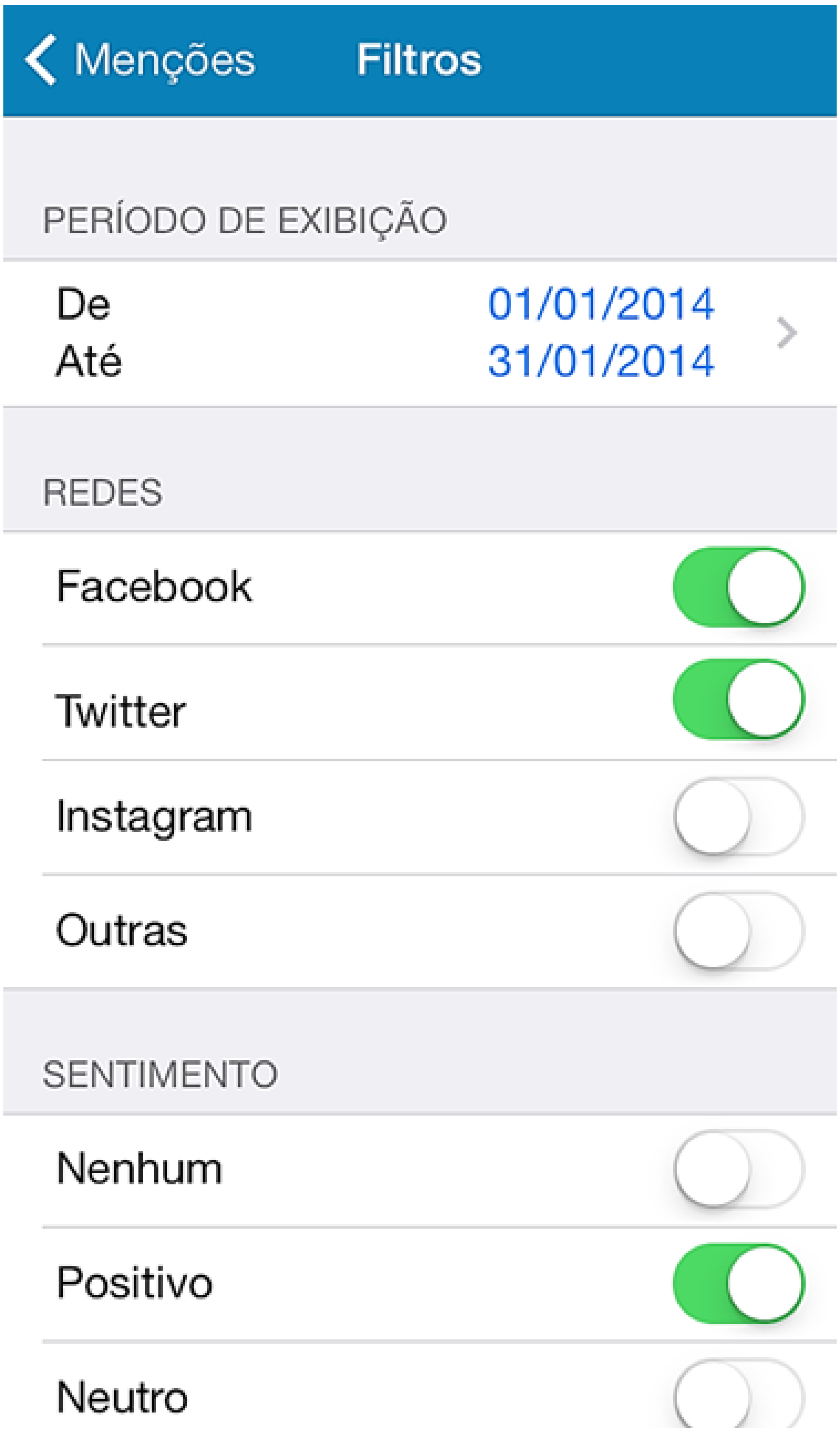